\documentclass[twocolumn,epjc3]{svjour3}          

\usepackage{graphicx}
\usepackage{dcolumn}
\usepackage{bm}
\usepackage{amsmath,multirow}
\usepackage{amssymb}
\usepackage{subfigure}
\usepackage{color}
\usepackage{verbatim} 

\RequirePackage[T1]{fontenc}

\smartqed  

\journalname{EPJC}

\begin{document}

\title{Impact of electric charges on chaos in magnetized Reissner-Nordstr\"{o}m spacetimes}


\author{
Daqi Yang$^{1,2}$ \and Wenfang Liu$^{1,2}$ \and
Xin Wu$^{1,2 \dag}$} 

\thankstext{e2}{e-mail: wuxin$\_$1134@sina.com}

\institute{School of Mathematics, Physics and Statistics, Shanghai
University of Engineering Science, Shanghai 201620, China
\label{addr1}
          \and Center of Application and Research of Computational Physics,
Shanghai University of Engineering Science, Shanghai 201620, China
\label{addr2}
}

\date{Received: date / Accepted: date}

\maketitle

\begin{abstract}

We consider the motion of test particles around a
Reissner-Nordstr\"{o}m black hole immersed into a strong external
magnetic field modifying the spacetime structure. When the
particles are neutral, their dynamics are nonintegrable because
the magnetic field acts as a gravitational effect, which destroys
the existence of a fourth motion constant in the
Reissner-Nordstr\"{o}m spacetime. A time-transformed explicit
symplectic integrator is used to show that the motion of neutral
particles can be chaotic under some circumstances. When test
particles have electric charges, their motions are subject to an
electromagnetic field surrounding the black hole as well as the
gravitational forces from the black hole and the magnetic field.
It is found that increasing both the magnetic field and the
particle energy or decreasing the particle angular momentum can
strengthen the degree of chaos regardless of whether the particles
are neutral or charged. The effect of varying the black hole
positive charge on the dynamical transition from order to chaos is
associated with the electric charges of particles. The dynamical
transition of neutral particles has no sensitive dependence on a
change of the black hole charge. An increase of the black hole
charge weakens the chaoticity of positive charged particles,
whereas enhances the chaoticity of negative charged particles.
With the magnitude of particle charge increasing, chaos always
gets stronger.

\end{abstract}



\section{Introduction}
\label{sec:intro}

The standard  general relativistic black hole solutions, such as
the Schwarzschild spacetime, Reissner-Nordstr\"{o}m (RN)
spacetime, Kerr spacetime and  Kerr-Newman spacetime, are highly
nonlinear. However, they are integrable. This integrability is due
to the existence of four constants of motion in these spacetimes.
In this case, the motion of  photons or test particles around
these black holes is regular and nonchaotic. The
regular motion of particles around black holes was treated in the
standard textbooks [1-3]. There are many integrable curved
spacetimes in modified or alternative gravity theories [4-9].

When these black holes are surrounded by extra sources, the
motions of  photons, neutral test particles or charged test
particles in the close vicinity of black hole horizons may be
nonintegrable. A magnetic field, as an important extra source
existing around a black hole, has been supported by observational
evidence [10]. It may arise from the dynamo mechanism in
collisionless plasma of an accretion disk around the central black
hole [3]. However, a strong magnetic field
around a supermassive black hole in the center of the Galaxy is
not relevant to an accretion disc [11]. Such a magnetic field
surrounding the black hole  is an asymptotically uniform
large-scale electromagnetic field with a complicated structure in
the vicinity of a magnetar at large distance from the black hole.

If a magnetic field in the vicinity of a black hole
has a small intensity satisfying the condition
$B\ll 10^{19}M_{\odot}/M$ Gauss, where $M_{\odot}$ and $M$ are the
masses of the Sun and black hole, it has a negligible effect on
the motion of neutral test particles. In other words, the
gravitational background is not influenced by the magnetic field,
and  the metric tensor of the black hole spacetime has no
modification. In spite of this, the magnetic field strongly
influences the motion of charged test particles if the ratio of
the particle charge to the particle mass is quite large. Even the
dynamics of  charged test particles is nonintegrable and probably
chaotic. The motion of charged test particles around the RN black
hole combined with an external asymptotically uniform magnetic
field can be chaotic [12]. There have been a large variety of
papers on the chaotic charged particle dynamics in combined black
hole gravitational and external magnetic fields
[13-30]. The authors of [16] showed that the
chaotic charged particle dynamics close to the black hole horizon
plays an important role in energy interchange between the
translational and oscillatory modes. It is not necessarily correct
that magnetic fields in the vicinity of black holes always lead to
the nonintegrability of charged particle motion around the black
holes.  The dynamics of charged particles near the Kerr-Newman
black hole immersed in an external magnetic field is integrable
due to the existence of the Carter constant as a fourth constant
of motion [31]. The regularity of the motion of
charged particles in the Kerr-Newman-de Sitter dyon spacetimes was
treated by Stuchl\'{i}k [32]. When a magnetic charge of the
Kerr-Newman black hole is included, the geodesic motion of charged
test particles is still integrable [33]. The integrability is also
suitable for charged particle motion around the Kerr-Newman black
hole with quintessence, cloud of strings and external
electromagnetic field [34].

When the strength of external magnetic field
reaches the upper limit of magnetic field $B= 10^{19}M_{\odot}/M$
Gauss, the magnetic field can significantly modify the black hole
spacetime structure [16,35,36]. In this case, the metric tensor
of the black hole spacetime needs an appropriate modification.
Such a magnetized black hole spacetime is likely nonintegrable
although the original nonmagnetized black hole spacetime is
integrable. In Melvin's magnetic universes, the Schwarzschild
black hole, RN black hole, and Kerr-Newman black hole, which were
derived from the coupled Einstein-Maxwell field equations by Ernst
[37], are not integrable. This nonintegrability
is due to the black hole immersed into the strong external
magnetic field modifying the spacetime structure and causing the
absence of the fourth motion constant. The result was shown by
finding the chaoticity of neutral test particles in the magnetized
Schwarzschild-Melvin spacetime [38,39]. The chaotic motion of
photons can give self-similar fractal structures to the shadows of
Schwarzschild-Melvin and Kerr-Melvin black holes [40,41]. For the
motion of charged test particles, not only the external fields
appear in the metrics, but also external electromagnetic fields
are included in the Hamiltonian systems associated with these
magnetized black hole metrics. Thus, the external magnetic fields
have typical effects on the motion of neutral and charged test
particles. Other properties of the RN-Melvin black hole solutions
were investigated in [42,43]. Note that the external magnetic
field in [12] is not included in the RN black hole spacetime, but
is added to the Hamiltonian system describing the motion of
charged test particles around the RN black hole. Recently, the
authors of [44] showed that the radii of the innermost stable
circular orbits for neutral and charged test particles around the
magnetized electric RN black hole could be strongly influenced by
the combined effect of black hole electric charge and magnetic
field. More recently, the tachyonic instability of RN-Melvin black
holes in Einstein-Maxwell-scalar theory was considered in Ref.
[45].

The self-force of the motion in combined
gravitational and magnetic fields plays an important role in
causing transitions from regular to chaotic motion (see e.g.
[29,30]). Besides the magnetic fields, other extra sources may
make  crucial contributions to the occurrence of chaos. Several
works [46-48] have shown that the quadrupolar deformations of
black hole masses are responsible for the existence of chaotic
dynamics of test particles in rotating black hole solutions of
Manko et al [49]. The axially symmetric deformation described by
the mass density parameter is necessary for the existence of
chaotic dynamics in the Zipoy-Voorhees metric [50]. Spin effects
of test particles in black hole spacetime backgrounds can induce
chaos [51]. The general relativistic Poynting-Robertson effect
shows a chaotic behavior of the geodesic motion of test particles
orbiting around the Kerr black hole [52,53].

The detection of the chaotical behavior needs very accurate
long-time determination of the trajectories. The motions of test
particles in many curved spacetimes can be described by
Hamiltonian systems. The most appropriate methods for solving the
Hamiltonian systems in the case of long-term integrations are
symplectic schemes (see, e.g., [54-57]). Such integrators preserve
the symplectic structure of Hamiltonian dynamics. Although they
are unlike the energy-preserving integrators [58-62] that can
exactly preserve energy for Hamiltonian systems, they have no
secular drifts in errors of first integrals of the Hamiltonian
systems. Because most of the Hamiltonians for curved spacetimes
are nonseparable to the position and momentum variables or cannot
be split into two explicitly integrable parts, explicit symplectic
integrations had been seldom applicable for these nonseparable
Hamiltonian problems. Of course,  implicit symplectic integration
schemes [14,48] or explicit and implicit combined symplectic
methods [62-65] are always available, and hence are
computationally expensive. Recently, explicit symplectic
integrators [12, 21, 22, 25, 26] have been made for nonseparable
Hamiltonian problems of geodesics in some curved spacetimes, such
as the Schwarzschild black hole spacetime. Their constructions are
based on the Hamiltonians split into more parts, where the flow of
each part can be integrated and represented in terms of explicit
functions of time. For some other curved spacetimes (e.g., the
Kerr black hole spacetime), their Hamiltonians have no such
splits. However, these  Hamiltonians have via appropriate time
transformations and thus amenable for explicit symplectic
integrations [23, 24, 27, 28, 66, 67].

In this paper we use an explicit symplectic integrator to
investigate the regular and chaotic dynamics of neutral and
charged test particles around the magnetized electric RN black
holes [42]. We first briefly introduce the magnetized electric RN
black hole metric. A Hamiltonian system for the motion of neutral
and charged test particles around the black holes is presented.
Next, we demonstrate how  several explicit symplectic methods are
constructed for this Hamiltonian system.  We then evaluate the
numerical performance of the proposed explicit symplectic methods
and find the method satisfying  a requirement for good long term
behaviour. Finally, we study the motions of neutral and charged
test particles. We particularly focus on the impact of varying the
black hole electric charge and the particle electric charge on a
dynamical transition from order to chaos. The effect of varying
the other parameters on the dynamical transition is also
considered. Explanations are given to the effects of the
parameters on the dynamics.

\section{Magnetized electric RN black hole}

At first, a magnetized electric RN black hole metric is
introduced. Then, the motion of a charged test particle around the
RN black hole surrounded by an asymptotically uniform magnetic
field is described in terms of a Hamiltonian formulation.

\subsection{Black hole metric}

In  the Boyer-Lindquist dimensionless coordinates $x^{\alpha}=(t,
r, \theta, \phi)$,  a magnetically charged RN black hole metric is
written in Refs. [37,42] as
\begin{eqnarray}
    ds^2 &=& g_{\alpha \beta}dx^\alpha dx^\beta \nonumber \\
&=& g_{tt}dt^2+2g_{t\phi} dt d\phi +g_{rr}dr^2 \nonumber \\
&& +g_{\theta\theta}d\theta^2 +g_{\phi\phi}d\phi^2; \\
g_{tt} &=& -fF+\frac{\omega^2}{F}r^2\sin^2\theta, ~~~~ g_{t\phi} =
-\frac{\omega}{F}r^2\sin^2\theta, \nonumber
\\ g_{rr} &=& \frac{F}{f}, ~~~~ g_{\theta\theta} = Fr^2, ~~~~
g_{\phi\phi} = \frac{1}{F}r^2\sin^2\theta. \nonumber
\end{eqnarray}
Here, $F$, $f$ and $\omega $ are functions of $r$ and $\theta$ as
follows:
\begin{eqnarray}
F &=& 1+\frac{1}{2}B^2(r^2\sin^2\theta+3Q^2\cos^2\theta) \nonumber \\
    &&+\frac{1}{16}B^4(r^2\sin^2\theta+Q^2\cos^2\theta)^2,\\
f &=& 1-\frac{2M}{r}+\frac{Q^2}{r^2},\\
\omega &=& -\frac{2QB}{r}+\frac{1}{2}QB^3r(1+f\cos^2\theta).
\end{eqnarray}
The speed of light $c$ and  the gravitational constant $G$ take
one geometric unit, $c=G=1$. $M$ is the mass of the black hole and
$Q$ denotes an electric charge of the black hole. $B$ stands for
the strength of an asymptotically uniform magnetic field in the
black hole vicinity. The magnetic field is included in the
spacetime geometry near the black hole because it is strong enough
to distort the spacetime geometry. It gives gravitational effects
rather than  the Lorentz force contributions to neutral test
particles. Of course, it must drastically affect the motion of
neutral particles. In this sense, the RN spacetime geometry is
 magnetized.

If $B=0$, the metric (1) corresponds to the RN black hole. If
$B\neq0$ and $Q=0$, the metric (1) is the Schwarzschild-Melvin
magnetic universe [37]. When the electric charge of the black hole
is nonzero, the metric (1) is not asymptotic to the static Melvin
metric [42]. Although the gravitational effect of the magnetic
field does not alter the even horizons of the RN black hole, it
causes the spacetime (1) not to be asymptotically flat.

The presence of the term $dtd\phi$ in the metric (1) is not due to
the black hole rotating, but arises from the global $SU(2, 1)$
symmetry group [42]. When  a Kaluza-Klein reduction of the
four-dimensional Einstein-Maxwell action is performed and the
vector fields are dualized to scalars in three dimensions, the
specific $SU(2, 1)$ transformation can generate the magnetized
solutions from nonmagnetized ones. If $B=0$ or $Q=0$, then
$\omega=0$ and the term $dtd\phi$ is absent. It is clear that the
term $dtd\phi$ directly comes from the contribution of the
magnetic field $B$ and the electric charge $Q$. In fact, the black
hole is not rotating at all.

If the test particle is charged, it suffers from the Coulomb force
and Lorentz force given by an external electromagnetic field. The
electromagnetic field is described by a four-vector potential with
two non-zero covariant components [42]
\begin{eqnarray}
A_\phi  &=&\frac{2}{B}-\frac{1}{F}[\frac{2}{B}
+\frac{B}{2}(r^{2}\sin^2\theta+3Q^2\cos^2\theta)], \\
A_t &=&-\frac{Q}{r}+\frac{3}{4}QB^2r(1+f\cos^2\theta)-\omega
A_\phi.
\end{eqnarray}
Besides the Coulomb force and Lorentz force, the gravity forces
from the magnetized RN black hole are given to the charged
particle.

\subsection{Hamiltonian system}

Now, let us consider the  particle with charge $q$ and mass $m$
moving near the RN black hole surrounded by the external magnetic
field $B$. The motion of charged particle can be described by a
Hamiltonian system
\begin{eqnarray}
    H &=& \frac{1}{2m}g^{\alpha\beta}(p_\alpha -qA_\alpha)(p_\beta-qA_\beta) \nonumber \\
    &=&\frac{f}{2mF}p_r^2+\frac{1}{2mFr^2}p_\theta ^2+H_1,
\end{eqnarray}
where  $H_1$ is a function of $r$ and $\theta$:
\begin{eqnarray}
   H_1 &=& \frac{g^{tt}}{2m}(p_t -qA_t)^2+\frac{g^{\phi\phi}}{2m}(p_\phi
   -qA_\phi)^2 \nonumber \\
   && + \frac{g^{t\phi}}{m}(p_t -qA_t)(p_\phi
   -qA_\phi).
\end{eqnarray}
The non-zero contravariant components of the metric (1) are
written as
\begin{eqnarray}
   g^{tt} &=& \frac{g_{\phi\phi}}{g_{tt}g_{\phi\phi}-g^2_{t\phi}}=-\frac{1}{fF},\nonumber \\
   g^{\phi\phi} &=& \frac{g_{tt}}{g_{tt}g_{\phi\phi}-g^2_{t\phi}}=\frac{fF^2-\omega^2r^2\sin^2\theta}{fFr^2\sin^2\theta},
\nonumber \\
g^{t\phi} &=&
-\frac{g_{t\phi}}{g_{tt}g_{\phi\phi}-g^2_{t\phi}}=-\frac{\omega}{fF}.
\nonumber
\end{eqnarray}
The external magnetic field in Eqs. (1) and (7) affects not only
the spacetime geometry but also the motion of charged particles.
The external magnetic field of [12] does not change the spacetime
geometries, but has a nonnegligible effect on the motion of
charged test particles in the gravitational backgrounds.

Because the Hamiltonian (7) does not explicitly depend on the
coordinate time $t$, the momentum $p_t$ is a motion constant
related to the particle energy $E$ with $E=-p_t$. This Hamiltonian
does not explicitly contain $\phi$ or is axially symmetric,
therefore, the momentum $p_\phi$ corresponds to the particle
conserved angular momentum $L=p_\phi$. The two constants satisfy
the relations
\begin{eqnarray}
\dot{t} &=& -\frac{g^{tt}}{m}(E +qA_t)+ \frac{g^{t\phi}}{m}(L
   -qA_\phi),\\
\dot{\phi} &=& \frac{g^{\phi\phi}}{m}(L
   -qA_\phi)- \frac{g^{t\phi}}{m}(E+qA_t).
\end{eqnarray}
Here, $\dot{t}$ and $\dot{\phi}$ as two components of the
4-velocity are derivatives of $t$ and $\phi$ with respect to the
proper time $\tau$. A third constant of motion is the conserved
Hamiltonian. For time-like geodesic orbit, this constant is
\begin{equation}
H=-\frac{m}{2}.
\end{equation}

For simplicity, dimensionless operations are used via scale
transformations to the related variables and parameters: $r\to
rM$, $t\to tM$, $\tau\to \tau M$, $Q\to QM$, $H \to mH$, $E \to
mE$, $p_r \to mp_r$, $p_\theta \to mMp_\theta$, $L \to mML$, $q
\to mq$, and $B \to B/M$. In this way, the two
mass factors $M$ and $m$ in Eqs. (1)-(11) are eliminated. In tis
case, Eq. (8) is rewritten as
\begin{eqnarray}
  H_1 &=& \frac{g^{tt}}{2}(E+qA_t)^2+\frac{g^{\phi\phi}}{2}(L
   -qA_\phi)^2 \nonumber \\
   && - g^{t\phi}(E+qA_t)(L-qA_\phi).
\end{eqnarray}

Besides the three constants given in Eqs. (9)-(11), no fourth
motion constant exists in the Hamiltonian system (7). The system
is nonintegrable even if the particle has no charge, namely, the
four-vector potential terms $A_t$ and $A_{\phi}$ are removed. The
nonintegrability of the motion of neutral particle is due to the
gravitational effect of the external magnetic field $B$ in the
gravitational background. Numerical techniques are convenient to
solve such a nonintegrable system.

\section{Construction of explicit symplectic integration algorithms}

A symplectic integrator that preserves the symplectic structure of
Hamiltonian dynamics is naturally a prior choice of an integrator
for solving the Hamiltonian problem (7). The Hamiltonian has no
separation of variables. It cannot be split into two parts with
analytical solutions as explicit functions of proper time, either.
In these cases, the construction of explicit symplectic
integrators seems to be impossible. Recently, the authors of
[12,21,22,25,26] showed that the Hamiltonians of some spacetimes
like the Schwarzschild black hole have more than two explicitly
interable splitting terms. Such splitting methods allow for the
construction of explicit symplectic integrators based on splitting
and composing. However, the Hamiltonian (7) has no such a direct
multi-part splitting. Introducing appropriate time transformations
to Hamiltonians of some other spacetimes such as the Kerr black
hole, the authors of [23,24,27,28,66,67] found that the obtained
time-transformed Hamiltonians are suitable for the application of
explicit symplectic methods. The Hamiltonian (7) belongs to Type 2
of the indirect splitting spacetimes in the latest paper [67].
Following this idea, we implement such algorithms for the
Hamiltonian (7).

\subsection{Algorithmic construction}

Setting the proper time $\tau $ as a new coordinate  $q_0=\tau $
and its corresponding momentum as $p_0=-H =1/2$, we extend the
phase space of the Hamiltonian (7) in the form
\begin{equation}
    \mathcal{H} =H +p_0.
\end{equation}
The extended phase space Hamiltonian is always identical to zero
for any proper time $\tau $, i.e., $\mathcal{H} =0$.

Following the time transformation method of  Mikkola [68], we take
a time transformation
\begin{equation}
    d\tau =g(r,\theta )d w,
\end{equation}
where $g$ is a time transformation function
\begin{equation}
    g(r,\theta )=F.
\end{equation}
We then have a time transformation Hamiltonian
\begin{eqnarray}
   K=g \mathcal{H}.
\end{eqnarray}
The time-transformed Hamiltonian is still equal to zero ($K=0$)
for any new time $w$. It has five splitting pieces
\begin{eqnarray}
    K =K_1+K_2+K_3+K_4+K_5,
\end{eqnarray}
where the five sub-Hamiltonians are expressed as
\begin{eqnarray}
    K_1 &=& g(H_1+p_0), \\
    K_2 &=& \frac{1}{2}p_r^2, \\
    K_3 &=& -\frac{1}{r}p_r^2, \\
    K_4 &=& \frac{p_\theta  ^2}{2r^2}, \\
    K_5 &=& \frac{Q^2}{2r^2}p_r^2.
\end{eqnarray}
The splitting of the time-transformed Hamiltonian $K$ is similar
to that of the Hamiltonian for the RN black hole with the external
magnetic field in Ref. [12].

It is clear that the five sub-Hamiltonians are explicitly,
analytically solvable. That is to say, their analytical solutions
are explicit functions of the new time $w$. Suppose that
$\mathcal{K}_1$, $\mathcal{K}_2$, $\mathcal{K}_3$, $\mathcal{K}_4$
and $\mathcal{K}_5$ correspond to the analytical solvers of the
five sub-Hamiltonians $K_1$, $K_2$, $K_3$, $K_4$ and $K_5$,
respectively. Let $h$ be a time step of the new time $w$. Two
first-order symplectic composing operators are defined as
\begin{eqnarray}
    \chi (h)&=& \mathcal{K}_1(h)\times\mathcal{K}_2(h)\times\mathcal{K}_3(h)\nonumber\\
        &&\times\mathcal{K}_4(h)\times\mathcal{K}_5(h), \\
    \chi^{\ast} (h) &=&
        \mathcal{K}_5(h)\times\mathcal{K}_4(h)\times\mathcal{K}_3(h)\nonumber\\
        &&\times\mathcal{K}_2(h)\times\mathcal{K}_1(h).
\end{eqnarray}

The two operators can symmetrically compose an explicit
second-order symplectic algorithm
\begin{eqnarray}
    S_2(h)=\chi^{\ast}(\frac{h}{2})\times\chi(\frac{h}{2}).
\end{eqnarray}
 A symmetric composition of three second-order methods easily raises a fourth-order method of Yoshida [69]
\begin{eqnarray}
    S_4(h)=S_2(\gamma h)\times S_2(\delta h)\times S_2(\gamma h),
\end{eqnarray}
where $\gamma =1/(1-\sqrt[3]{2}) $ and $\delta =1-2\gamma$. There
is an optimized fourth-order partition Runge-Kutta (PRK) explicit
symplectic integrator [70]:
\begin{eqnarray}\nonumber
    PRK_64(h) &=& \chi^{\ast} (\alpha _{12} h)\times \chi(\alpha
    _{11}
    h)\times \cdots \\
    && \times \chi^{\ast} (\alpha _2 h) \times \chi(\alpha _1 h),
\end{eqnarray}
The time coefficients of the algorithm are listed in Ref. [25] as
follows:
\begin{eqnarray}
    \nonumber
    &&\alpha _1=\alpha _{12}= 0.079203696431196,   \\ \nonumber
    &&\alpha _2=\alpha _{11}= 0.130311410182166,   \\ \nonumber
    &&\alpha _3=\alpha _{10}= 0.222861495867608,    \\ \nonumber
    &&\alpha _4=\alpha _9=-0.366713269047426,    \\ \nonumber
    &&\alpha _5=\alpha _8= 0.324648188689706,   \\ \nonumber
    &&\alpha _6=\alpha _7= 0.109688477876750.   \nonumber
\end{eqnarray}

Eqs. (25)-(27) are the time-transformed explicit symplectic
methods designed for the Hamiltonian (7). In these methods, fixed
time steps are used for the new time $w$, while variant time steps
may be considered for the proper time $\tau$. The authors of [67]
gave the time transformation function and Hamiltonian splitting
form like those of Eqs. (15) and (17) to the Hamiltonian (13) with
$A_t=A_\phi=0$, but did not numerically test the established
time-transformed explicit symplectic methods.

\subsection{Evaluation of the algorithms}

The parameters are the magnetic field strength $B=6\times
10^{-4}$, black hole charge $Q=0.3$, particle charge $q=0.5$,
particle angular momentum $L=4.8$, and particle energy $E=0.997$.
The initial conditions are $r=30$, $\theta=\pi/2$ and $p_r=0$. The
initial value of $p_{\theta}>0$ should satisfy Eq. (16).

Taking the time step $h=1$, we plot Fig. 1, which shows accuracies
of the Hamiltonian $K$ yielded by the three methods $S_2$, $S_4$
and $PRK_64$. The errors of $K$ remain stable for $S_2$ and $S_4$,
but have a secular drift for $PRK_64$. The errors for $S_4$ are
about three orders of magnitude smaller than those for $S_2$, but
larger than those for $PRK_64$. The secular error drift for
$PRK_64$ is due to the rapid growth of roundoff errors. It is
absent when the time step increases to $h=3.5$. The errors of
$PRK_64$ for the larger time step $h=3.5$ are  approximately the
same as those of $S_4$ for the smaller time step $h=1$. The
algorithm $S_4$ with $h=1$ has an advantage over $PRK_64$ with
$h=3.5$ in computational efficiency, as is shown in Table 1.

The relation between the new time $w$ and the proper time $\tau$
in Fig. 2 shows that the two times are almost the same. It means
that the time step is fixed for the new time $w$, but the proper
time steps appropriately remain invariant for the proper time
$\tau$.

Based on the accuracy and efficiency, the method  $S_4$ with $h=1$
is used to survey the orbital dynamics in later discussions.

\begin{table*}[htbp]
    \centering \caption{
    CPU times [units: minute($'$), second(")] for the three algorithms with two step sizes $h$.
    The initial separations of Orbit 1 and Orbit 2 are $r=30$ and $r=60$,
    respectively; the parameters and other initial conditions are the same as those of Fig.
    1.
    }
     \label{Tab1}
    \begin{tabular}{lcccccccccccc}
    \hline   & $S_2(h=1)$   & $S_4(h=1)$   & $S_4(h=3.5)$ &  $PRK_64(h=3.5)$ &  $PRK_64(h=1)$  \\
    \hline
    Orbit 1 & 0$'$25.08"     & 1$'$28.95"    & 0$'$27.91"     &  1$'$44.77"         &  5$'$24.11" \\
    \hline
    Orbit 2 & 0$'$24.59"     & 1$'$30.19"    & 0$'$28.27"     &  1$'$43.84"         &  5$'$30.16"  \\
    \hline
    \end{tabular}
\end{table*}

\begin{figure}[hpb]
\center{
\includegraphics[scale=0.25]{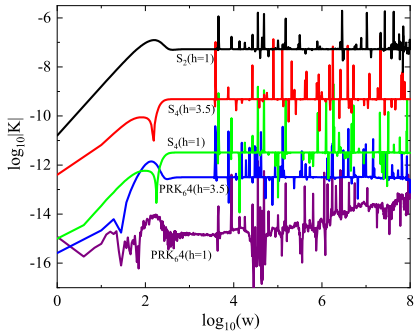}
\caption{Accuracies of the Hamiltonian $K$ for the three explicit
symplectic  integrators $S_2$, $S_4$ and $PRK_64$ with two step
sizes $h=1$ and  $h=3.5$. For the motion of a charged particle
with $q=0.5$, the other parameters are $E=0.997$, $L=4.8$,
$B=6\times 10^{-4}$ and $Q=0.3$; the initial conditions are
$r=30$, $p_r=0$, $\theta = \pi /2$, and $p_\theta >0$ given by Eq.
(16). A secular drift occurs in the energy errors for $PRK_64$
with $h=1$, whereas does not occur for $PRK_64$ with $h=3.5$.
    }
    \label{Fig1}}
\end{figure}

\begin{figure}[hptb]
    \center{
    \includegraphics[scale=0.25]{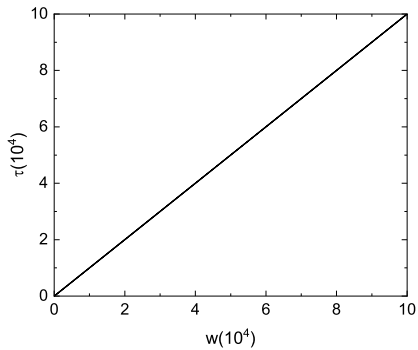}
    \caption{
Relation between the proper time $\tau $ and the new time $w$. The
tested orbit is that of Fig. 1. The slope being 1 shows that the
two times are almost the same. }
    \label{Fig2}}
\end{figure}

\section{Orbital dynamics}

Using several chaotic indicators, we focus on the dynamics of
neutral or charged particles moving around the magnetized electric
RN black hole. The effects of varying one or two parameters on a
transition from order to chaos are also considered.

\subsection{Dynamics of neutral particles}

For the motion of neutral particles with $q=0$, the terms $A_t$
and $A_{\phi}$ are dropped in Eqs. (8)-(10). The parameters are
taken as $B=6\times 10^{-4}$, $L=4.8$, $Q=0.3$ and $E=0.997$. The
initial conditions are $\theta = \pi /2$ and $p_r = 0$. The
initial radii have various choices, and the initial values
$p_\theta >0$ are calculated by Eq. (13). Fig. 3(a) relates to the
Poincar\'{e} map at the plane $\theta = \pi /2$ with $p_\theta
>0$, which represents intersections $(r, p_r)$ of the
particle trajectories with the surface of section. Orbit 1 with
the initial radius $r=30$ is chaotic because the intersections are
randomly distributed points in an area. However, Orbit 2 with the
initial radius $r=60$ has intersection points forming a closed
curve. As a result, the motion is regular.

In addition to the Poincar\'{e} section method, the largest
Lyapunov exponent for measuring the average deviation between two
adjacent orbits is often used to distinguish between ordered and
chaotic motions. It is defined in [71] by
\begin{eqnarray}
    \lambda =\lim_{w \to \infty}  \frac{1}{w} \ln \frac{d(w)}{d(0)} ,
\end{eqnarray}
where $d(w)$ and $d(0)$ are  the distances between two adjacent
orbits at the new time $w$ and the starting time, respectively.
When the integration time reaches $w=1\times 10^8$ in Fig. 3(b),
the Lyapunov exponent $\lambda$ of bounded orbit 1 tending to  a
stabilizing value $10^{-3.761}$ indicates the chaoticity of Orbit
1. The Lyapunov exponent $\lambda$ of bounded orbit 2 tending to
zero indicates the regularity of Orbit 2.

The fast Lyapunov indicator (FLI) [72] is a faster and more
sensitive tool to detect chaos from order than the largest
Lyapunov exponent. It is defined in [73] by
\begin{eqnarray}
    FLI =\log _{10} \frac{d(w)}{d(0)} .
\end{eqnarray}
Different growth rates of the deviation vectors are used to
distinguish between the regular and chaotic two cases. The the
exponential growth  of the FLI of Orbit 1 with time shows that of
chaotic orbit 1, whereas the algebraical  growth  of the FLI of
Orbit 2 with time $\log _{10} w$ describes the characteristic of
regular orbit 2 in Fig. 3(c).

The FLI is convenient to find chaos by scanning one or two
parameter spaces. Taking the initial conditions $\theta=\pi/2$,
$p_r=0$, $r=30$ and the parameters $E=0.998$, $L=4.5$, we plot the
dependence of FLIs on the parameters $Q$ and $B$ in Fig. 4(a).
When a pair of the values $Q$ and $B$ are given, each of the FLIs
is obtained after the integration time $w=1\times 10^6$. The FLIs
not more than 5 correspond to ordered orbits, while those larger
than 5 show chaotic orbits. The dynamical transition is
sensitively dependent on varying the magnetic field strength $B$.
As $B$ increases, the degree of chaos is typically enhanced.
However, the dynamical transition to chaos exhibits no sensitive
dependence on a change of the black hole charge $Q$. This result
is also shown in Fig. 4 (b) and (c). In addition, the occurrence
of chaos is easier when the energy $E$ increases or the angular
momentum $L$ decreases.

The Poincar\'{e} sections in Fig. 5 are used to check the results
of Fig. 4.  More orbits become chaotic and the strength of chaos
increases when the magnetic field strength increases from
$B=5\times 10^{-5}$ in Fig. 5 (a) to $B=3.5\times 10^{-4}$ in Fig.
5 (b), and to $B=6.5\times 10^{-4}$ in Fig. 5 (c). If a larger
magnetic field $B=4\times 10^{-4}$ is given, an increase of $Q$
seems to have no typical effect on the dynamical transition from
order to chaos in  Fig. 5 (d)-(f). When a smaller magnetic field
$B=5.5\times 10^{-5}$ is fixed,  the changes of $Q$  in  Fig. 5
(g)-(i) seem to exert no explicit influences on the dynamics of
neutral particles. The method of Poincar\'{e} sections in Fig. 6
is used to check the results of Fig. 4 regarding  the influence of
varying the particle energy $E$ and angular momentum $L$ on chaos.
The increase of $E$ in Fig. 6 (a)-(c) leads to enhancing the
extent of chaos, while that of $L$ in Fig. 6 (a)-(c) results in
weakening the chaoticity of neutral particles. The results support
those of Fig. 4.

In a word, the methods of FLIs and Poincar\'{e} sections give the
consistent results regarding the effects of varying one or two
parameters on the dynamical transition. The
dependence of chaos on different parameters behaves well in the
spirit of KAM (Kolmogorov-Arnold-Moser) theorem that is crucial
for the transitions between the chaos and regularity. The
regularity has to be related to the local minima of the effective
potential of the motion in the equatorial plane $\theta=\pi/2$. It
is because the local minima correspond to the existence of stable
circular orbits. The values in the vicinity of the local minima of
the effective potential correspond to regular KAM tori. See Ref.
[17] for more details on the minima of the effective potential.
The astrophysical origin and role of these minima were discussed
in [74]. As the parameters gradually increase, the KAM tori are
twisted and some tori are destroyed. The absence of some tori
brings the possibility for the occurrence of chaos. Of course, the
fundamental reason for the chaoticity of neutral particles is that
the magnetic field as gravitational effects causes the
non-integrability of the spacetime geometry.

\subsection{Dynamics of charged particles}

For the motion of charged particles with $q\neq0$, the terms $A_t$
and $A_{\phi}$ are present in Eqs. (8)-(10). In this case, not
only the Coulomb force and Lorentz force from the external
electromagnetic field but also the gravitational forces from the
magnetized black hole affect the motion of charged particles. Even
the Lorentz force has an important contribution to the motion of
charged particles around the black holes.

\subsubsection{Particles with positive charges}

For $q=0.5$, the initial conditions and other parameters of Fig. 7
(a)-(c) for using the FLIs to scan the two-dimensional parameters
correspond to those of Fig. 4 (a)-(c), respectively. The degree of
chaos is still strengthened when $B$ and $E$ increase or $L$
decreases in Fig. 7 (a)-(c), as is in Fig. 4 (a)-(c).

The FLIs with respect to the two parameters in Fig. 7 show that
the increase of $Q$ suppresses the occurrence of chaos. The
increase of $Q$ weakening and suppressing the chaoticity of
positive charged particles is also shown by the method of
Poincar\'{e} sections in Fig. 8.

Unlike the increase of the black hole charge $Q$,  the increase of
the particle positive charge $q$ can easily induce the occurrence
of chaos. This result can be described clearly by the methods of
FLIs and Poincar\'{e} sections in Fig. 9.

\subsubsection{Particles with negative charges}

Now, let us consider the motion of  negative charged particles
with $q<0$.

Seen from the FLIs with respect to varying the black hole charge
$Q$ in Fig. 10(a), chaos becomes stronger as $Q$ increases for the
particle having an appropriate negative charge. This result is
also supported by the method of Poincar\'{e} sections in Fig. 10
(b)-(d), which take three different values of $Q=$0.2, 0.5 and
0.8, and give $q$ the same value $-0.5$.

When the black hole charge $Q$ is given an appropriate value, the
FLIs in Fig. 10(a) show that a larger absolute value of the
particle negative charge $q$ brings stronger chaos. For $Q=0.5$,
$q=-0.15$, $-0.5$, $-0.8$ in Fig. 11 correspond to the degree of
chaos from weak to strong, as shown through the method of
Poincar\'{e} sections.

\subsection{Theoretical analysis and explanations }

The comparison among Figs. 4, 7 and 10(a) shows that the increase
of the black hole charge $Q$ exerts different influences on the
dynamical transition from order to chaos for neutral, positive
charged and negative charged particles.  For neutral particles, a
change of the black hole charge $Q$ does not sensitively cause the
dynamical transition from order to chaos under some circumstances.
When the black hole charge $Q$ increases, the degree of chaos is
weakened for positive charged particles, whereas strengthened for
negative charged particles. An increase of the magnitude of
particle charge $q$ can easily induce the occurrence of chaos
regardless of whether the particle charges are positive or
negative. As the magnetic field and the particle energy increase
or the particle angular momentum decreases, chaos is always
stronger for any one of the three types of test particles. In what
follows, we analytically interpret the effects of varying one or
two parameters on the dynamical transition from order to chaos.

For simplicity, the Hamiltonian $K$ (17) for the description of
the equatorial zero velocity motions of  neutral or charged
particles is considered, where $\theta=\pi/2$ and
$p_r=p_\theta=0$. In this case,  attractive forces balance
repulsive forces. In addition, the dynamics of the Hamiltonian $K$
is that of the sub-Hamiltonian $K_1$ (18). Considering $r\gg 2$,
we expand $K_1$ as follows:
\begin{eqnarray}
   K_1 &\approx&
   \frac{1}{2}\left(1-E^2+B^2L^2-BqL\right)-\frac{E^2}{r}\left(1+\frac{2}{r}\right)
   \nonumber \\
&&
+\frac{1}{4}B^2r^2+\frac{1}{8}B^2q^2r^2 +\frac{2}{r}BQEL+\frac{1}{r}EQq \nonumber \\
&& +\frac{L^2}{2r^2}+\frac{1}{2r^2}E^2Q^2 +\frac{4}{r^2}BQEL
+\frac{1}{2r^2}q^2Q^2
  \nonumber \\
&& +\frac{2}{r^2}EqQ.
\end{eqnarray}
The second term describes that the black hole gives an attractive
force to a particle.

For the motion of a neutral particle with $q=0$, the third term
$B^2r^2/4$ acts as an attractive force from the magnetic field.
The terms with repulsive force contributions to the particle are
the fifth term $2BQEL/r$, seventh term $L^2/(2r^2)$, eighth term
$E^2Q^2/(2r^2)$, and ninth term $4BQEL/r^2$. Because $E\sim1$
($E<1$), $L>3$, $0.1< Q\leq1$, and $0\leq B\ll1$, the eighth term
is more important than the fifth, ninth terms, but is denominated
by the seventh term. An increase of the energy $E$ or the magnetic
field $B$ leads to that of attractive forces, and therefore chaos
is easily induced under some circumstances. As the angular
momentum $L$ increases, the repulsive force increases, and then
the degree of chaos is weakened. With the black hole charge $Q$
increasing, the eighth term exerts a small influence on the
particle motion compared with the seventh term. This fact is why
the dynamical transition of neutral particles from order to chaos
does not sensitively depend on a change of the black hole charge.

When the particle positive charge $q$ with $q>0$ increases, the
Lorentz force  as an attractive force from the magnetic field of
the fourth term $B^2q^2r^2/8$ also increases. Therefore, chaos
occurs easily. If $Q$ increases for a given positive charge $q$,
the Coulomb force as a repulsive force from the sixth term $EQq/r$
increases, and is larger than the repulsive forces from the eighth
term $E^2Q^2/(2r^2)$, tenth term $q^2Q^2/(2r^2)$ and eleventh term
$2EqQ/r^2$. As a result, chaos becomes weaker.

When the magnitude of particle negative charge $q$ with $q<0$
increases, the Coulomb forces as the attractive forces from the
sixth, eleventh terms increase. Thus, chaos becomes stronger.
Clearly, this result is also suitable for the increase of the
black hole positive charge $Q$ for a given negative charge $q$.

\section{Conclusions}

The Hamiltonian for describing the motion of neutral, or charged
particles around the magnetized RN black holes cannot be split
into several parts, which have analytical solutions as explicit
functions of time. However, there are five explicitly integrable
splitting pieces through an appropriate time transformation to the
Hamiltonian. In this way,  explicit symplectic integrators can be
designed for the time-transformed Hamiltonian. These  symplectic
methods perform good numerical performance in long-term stabilized
behavior of energy errors for suitable choices of new time steps.
One of the integrators with the best performance is used to
provide some insight into the dynamics of particles.

The dynamics of neutral particles around magnetized RN black holes
is nonintegrable. This nonintegrability is
because the external magnetic field reaches the upper limit of
magnetic field  modifying the spacetime structure, and acts as a
gravitational effect destroying the integrability of the RN
spacetime. It can be chaotic under some circumstances. With the
magnetic field and the particle energy increasing or the particle
angular momentum decreasing, chaos is easily induced. This result
is also suitable for the motion of charged particles.

The effect of varying the black hole positive charge on the
dynamical transition from order to chaos is dependent on the
electric charges of test particles. A change of the black hole
charge does not sensitively affect the dynamical transition of
neutral particles. An increase of the black hole charge leads to
weakening the chaoticity of positive charged particles, but to
enhancing the chaoticity of negative charged particles. As the
magnitude of particle charge increases, chaos always gets stronger
regardless of whether the particle charges are positive or
negative.

\textbf{Acknowledgements}: The authors are very grateful to a
referee for valuable comments and suggestions. This research has
been supported by the National Natural Science Foundation of China
(Grant No. 11973020), and the Natural Science Foundation of
Guangxi (Grant No. 2019GXNSFDA245019).

\begin{figure*}[hptb]
    \center{
    \includegraphics[scale=0.2]{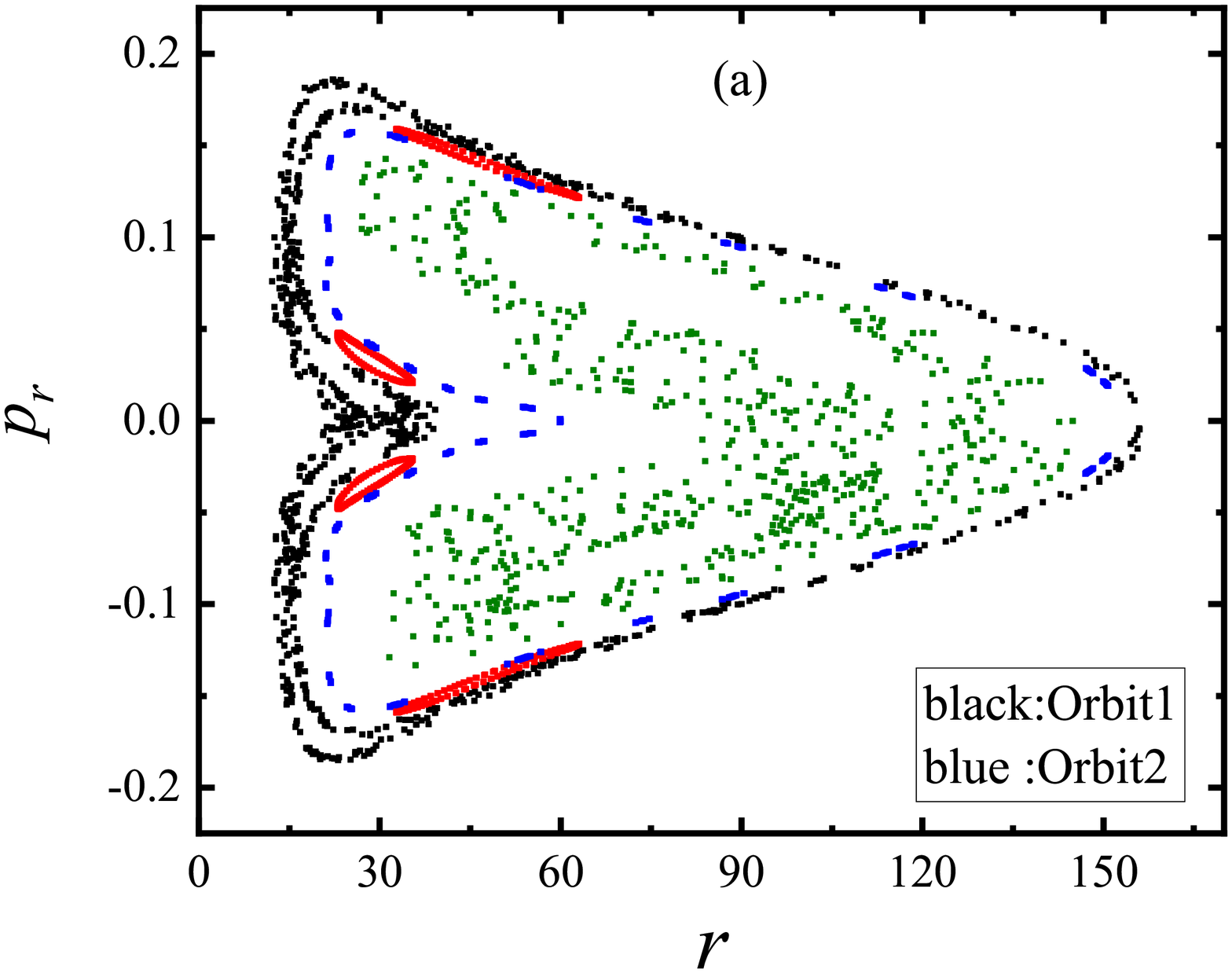}
    \includegraphics[scale=0.2]{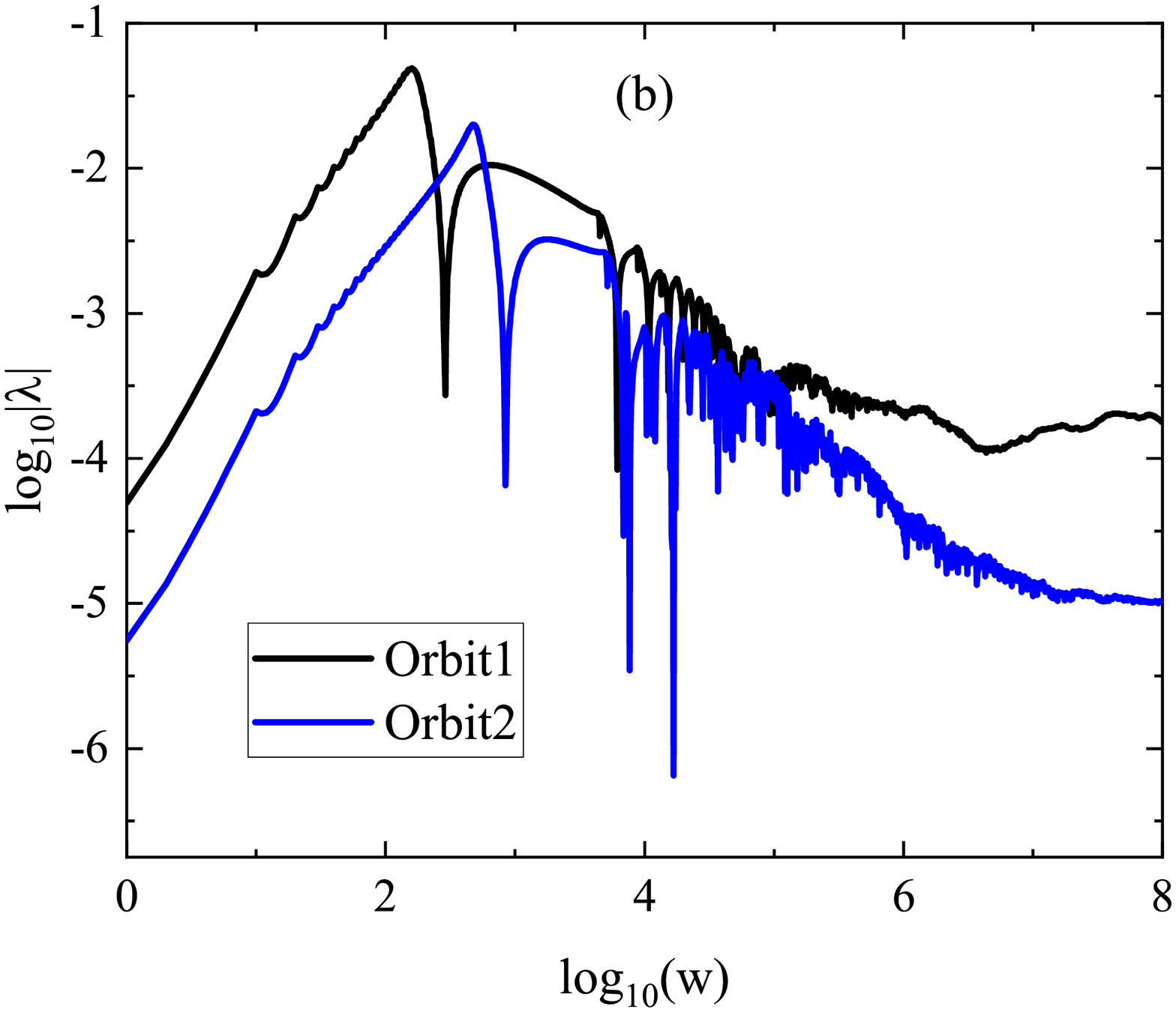}
    \includegraphics[scale=0.2]{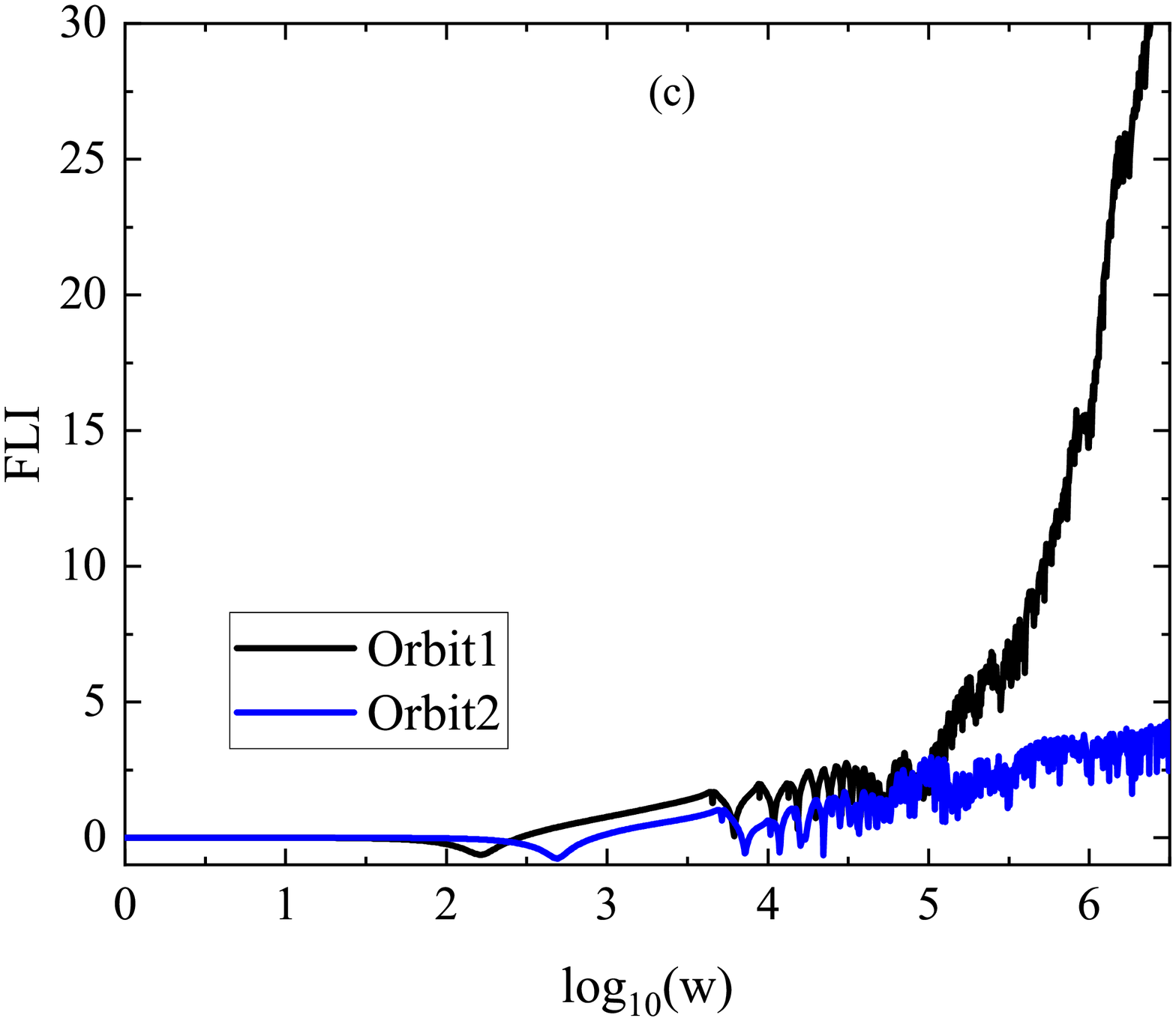}
    \caption{
(a) Poincar\'{e} sections at the plane $\theta = \pi /2$ with
$p_\theta >0$ for the motions of neutral particles with $q=0$. The
other parameters are those of Fig. 1. Orbit 1 with the initial
separation  $r=30$ is chaotic, but Orbit 2 with the initial
separation  $r=60$ is regular. These results are supported by the
largest Lyapunov exponents in panel (b) and fast Lyapunov
indicators (FLIs) in panel (c).
    }
     \label{Fig3}}
\end{figure*}

\begin{figure*}[hptb]
    \center{
    \includegraphics[scale=0.2]{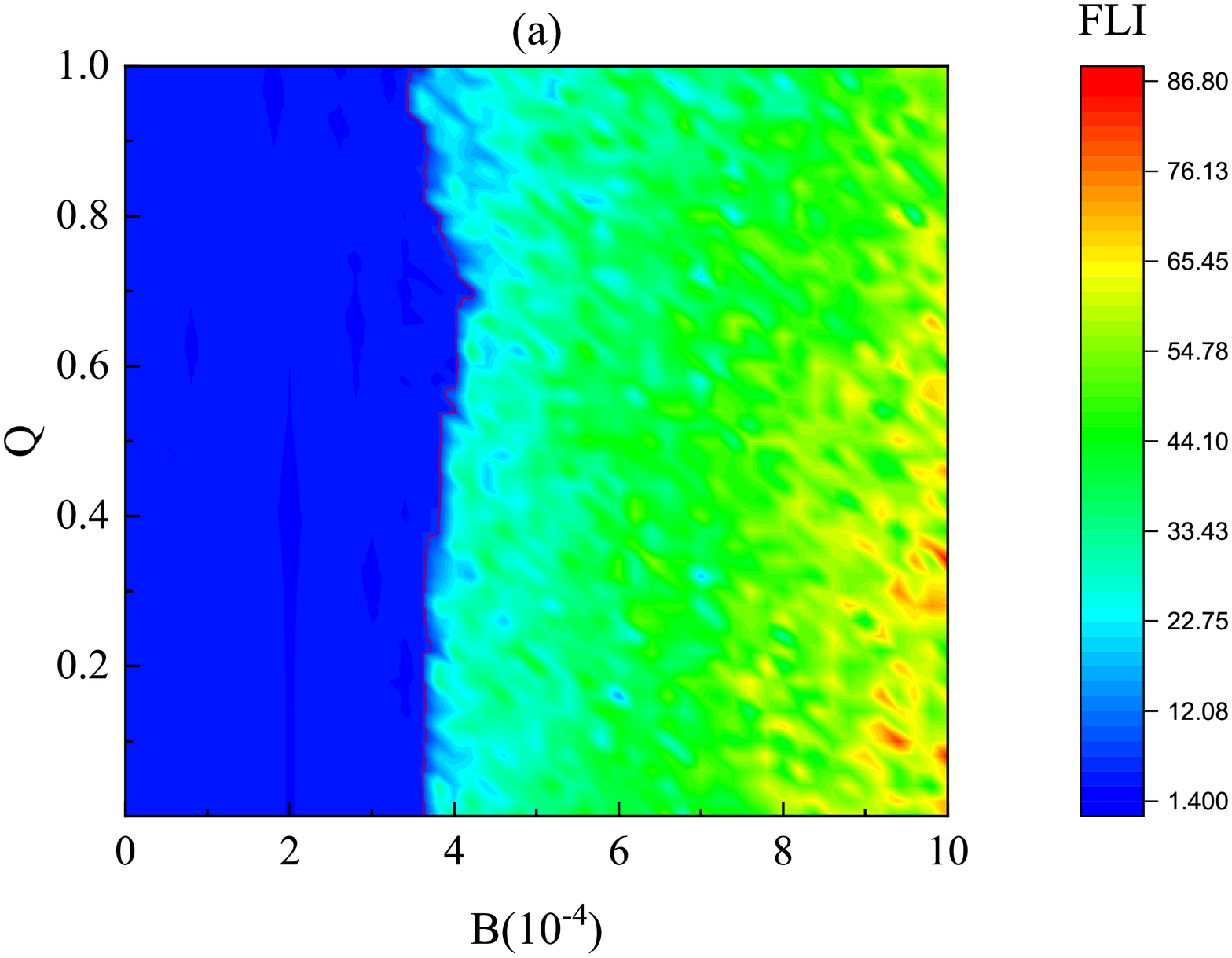}
    \includegraphics[scale=0.2]{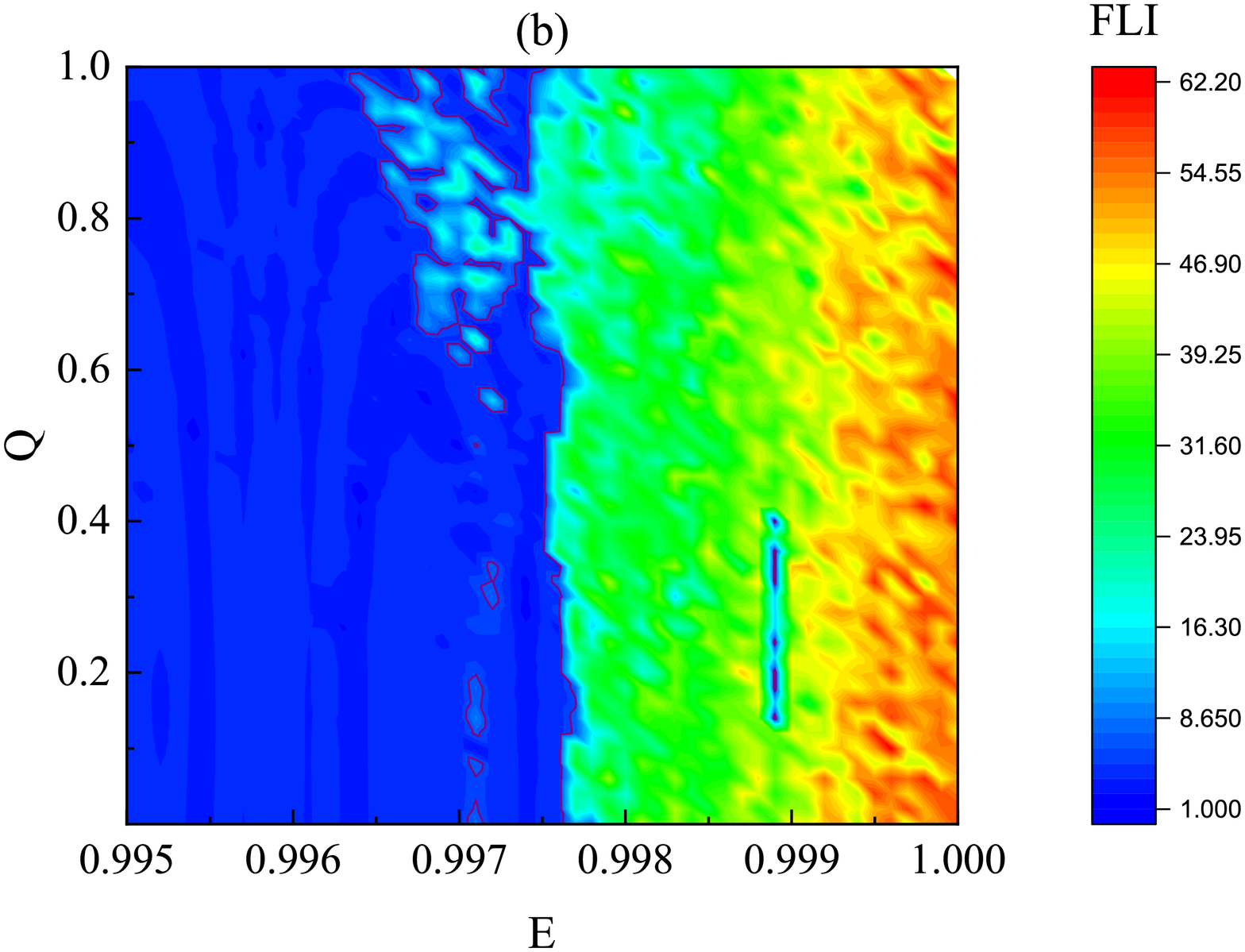}
    \includegraphics[scale=0.2]{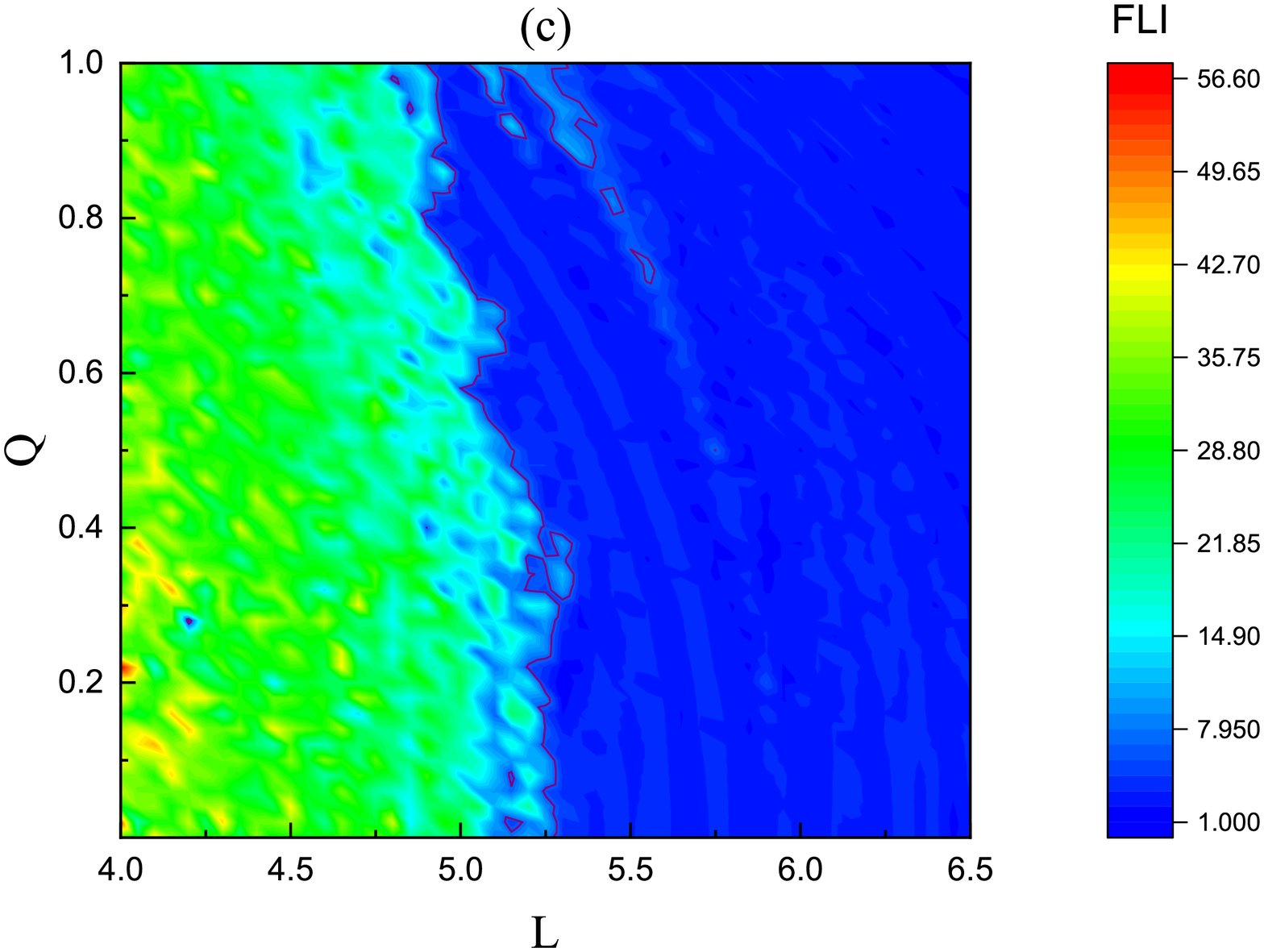}
    \caption{Dependence of FLIs on  two parameters. Each of the FLI values
    is obtained after the integration time $w=10^6$. The
    FLIs$\leq$5 show the regularity of bounded orbits, but the
    FLIs$>$5 describe the chaoticity of bounded orbits.
     (a) The two parameters are the magnetic field $B$ and
the black hole charge $Q$, and the other parameters are $q=0$,
$E=0.998$, and $L=4.5$; the initial separation is $r=30$. (b) The
two parameters are the particle energy $E$ and the black hole
charge $Q$, and the other parameters are $q=0$, $L=4.7$, and $B=5
\times 10^{-4}$; the initial separation is $r=30$. (c) The two
parameters are the particle angular momentum $L$ and the black
hole charge $Q$, and the other parameters are $q=0$, $E=0.998$ and
$B=4.5 \times 10^{-4}$; the initial separation is $r=45$. The
three panels clearly show that the degree of chaos increase with
the increase of $B$ and $E$ or the decrease of $L$. However, a
change of the black hole charge $Q$ has no explicit effect on the
dynamical transition from order to chaos.
    }
     \label{Fig4}}
\end{figure*}

\begin{figure*}[htbp]
\center{
\includegraphics[scale=0.2]{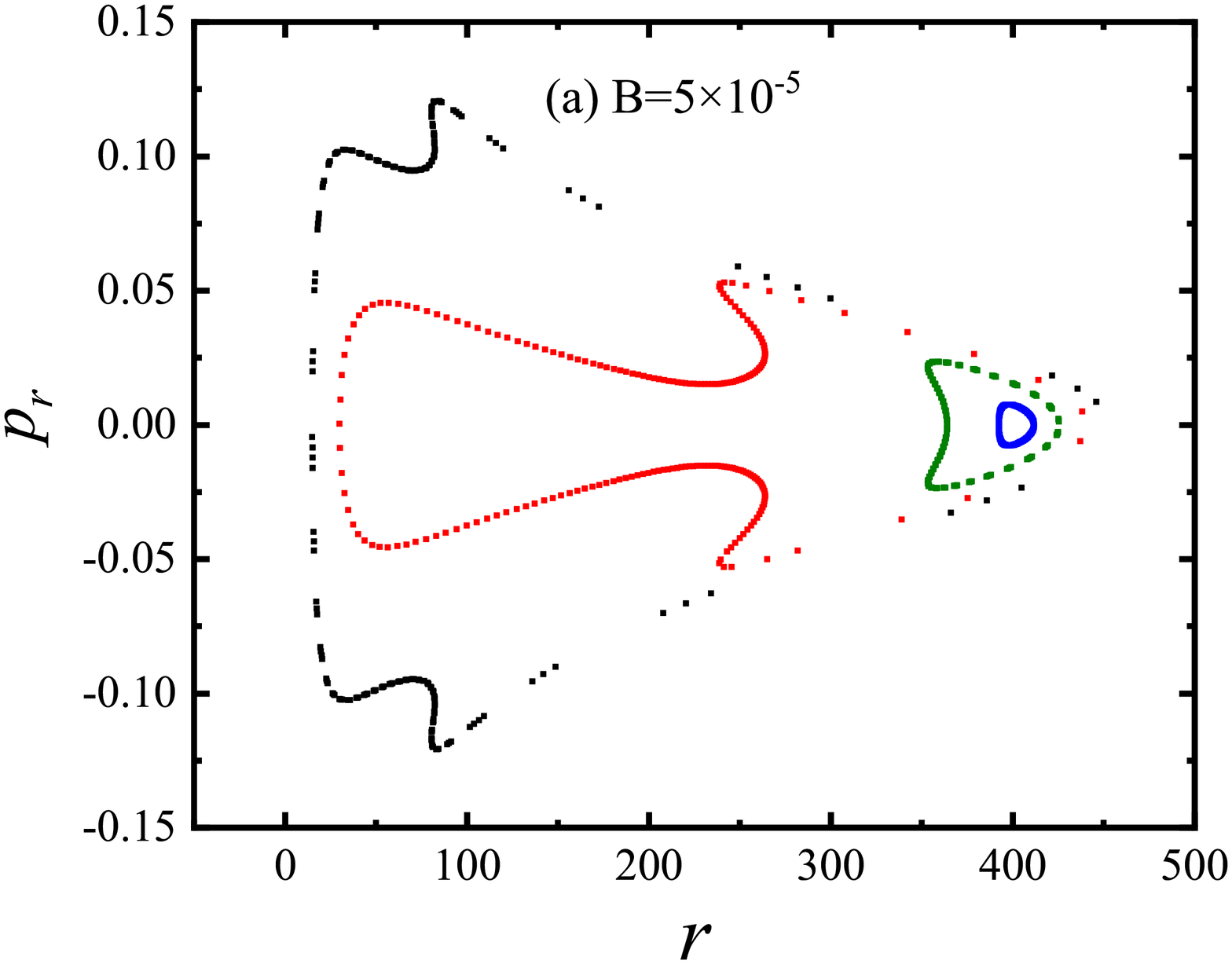}
\includegraphics[scale=0.2]{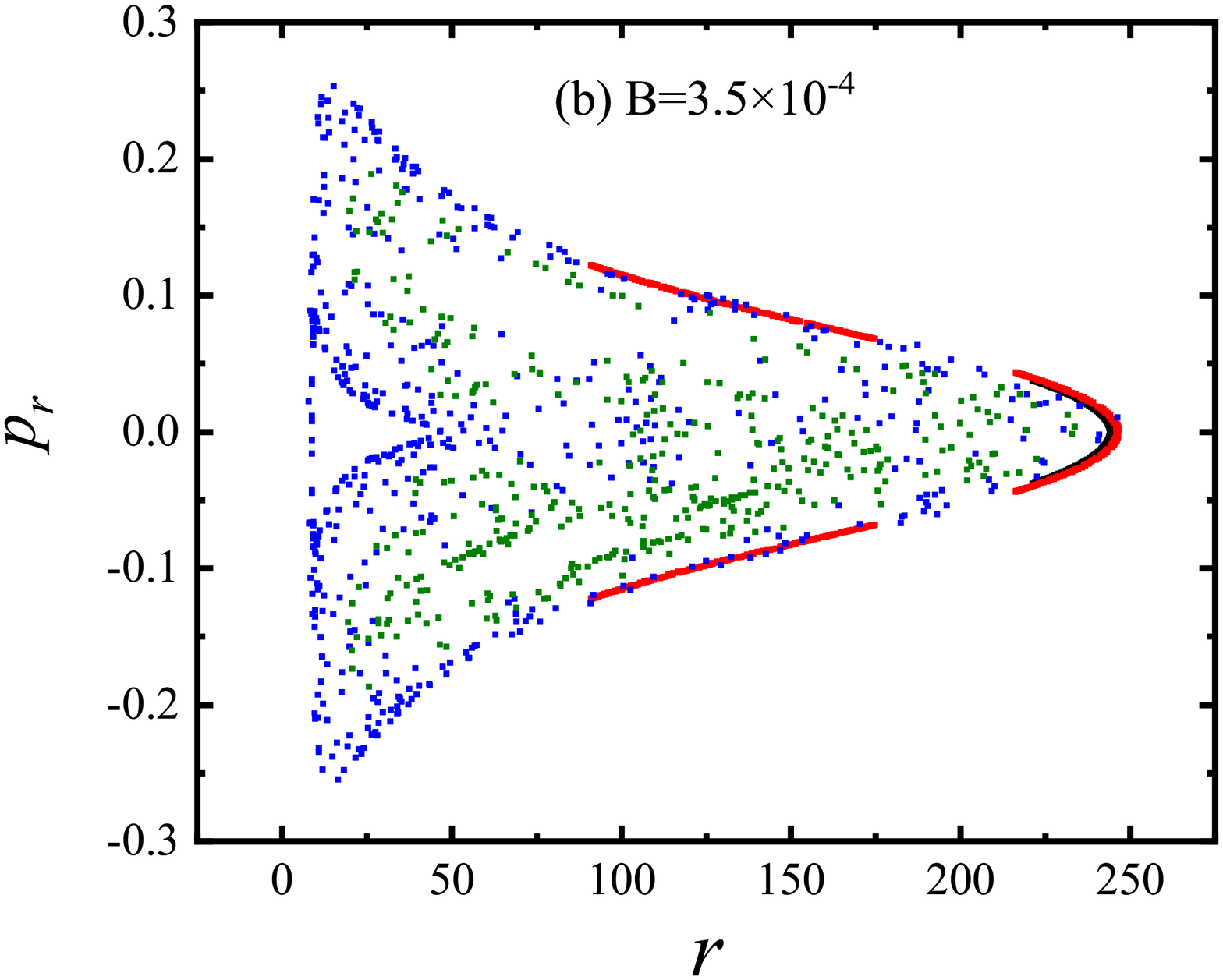}
\includegraphics[scale=0.2]{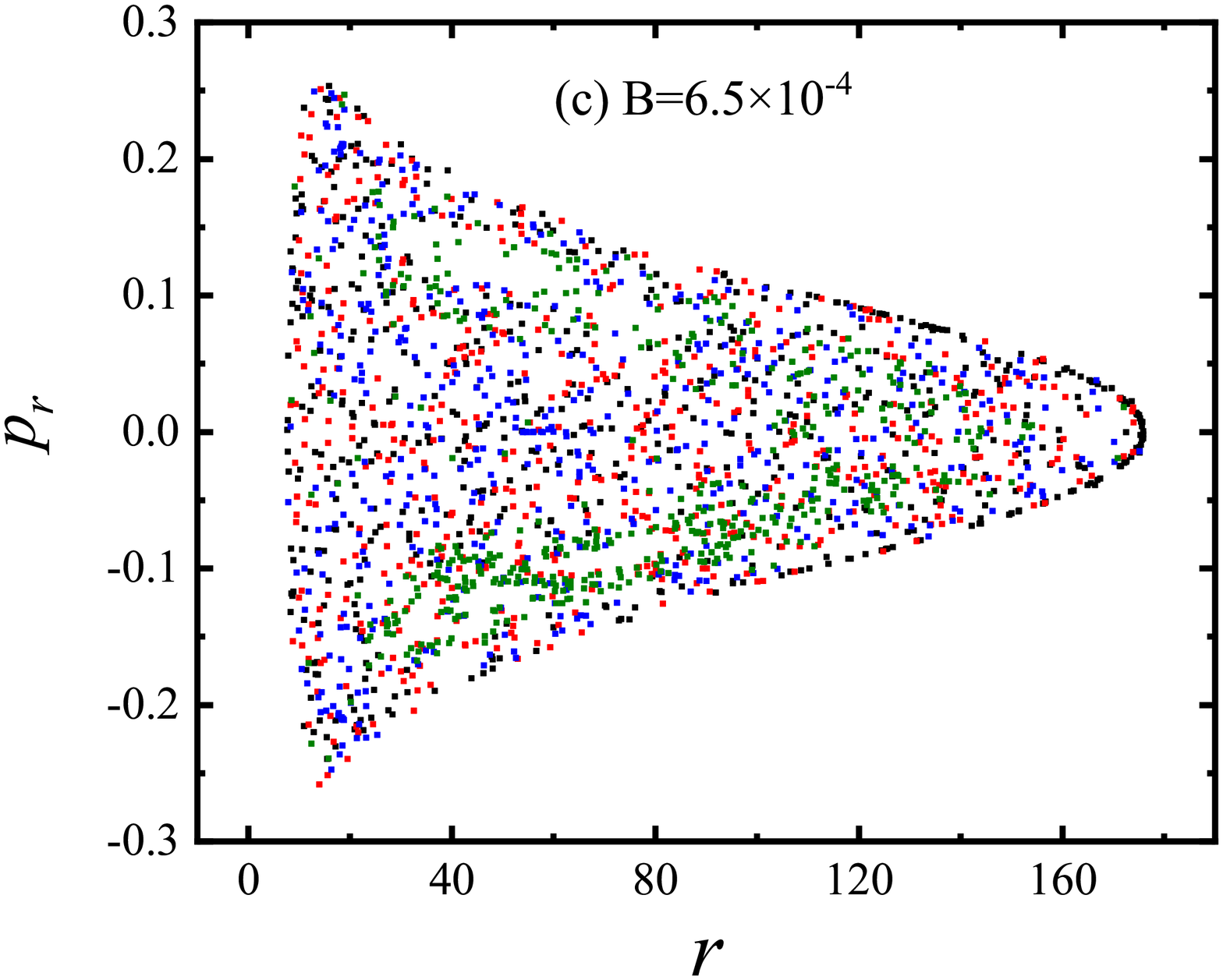}
\includegraphics[scale=0.2]{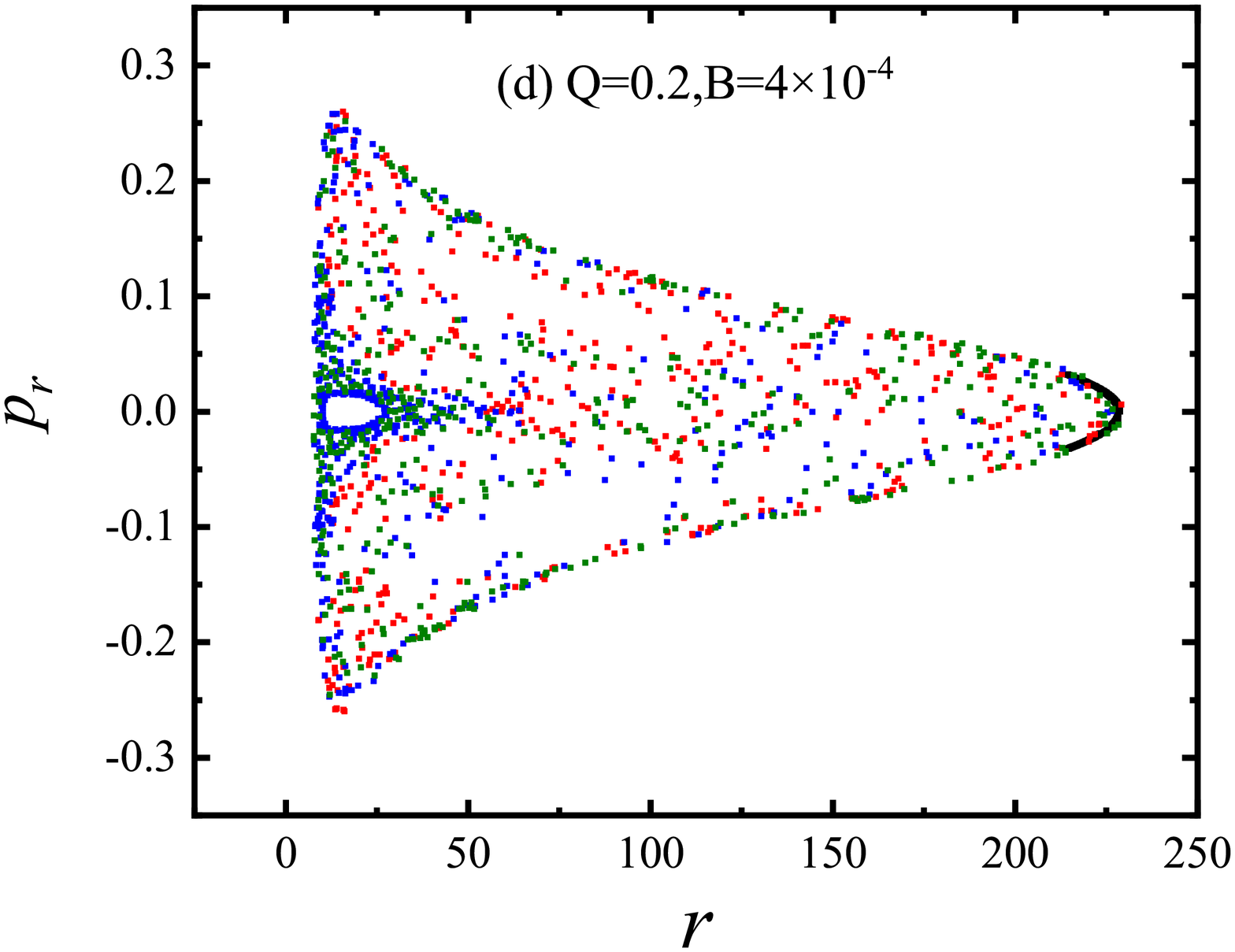}
\includegraphics[scale=0.2]{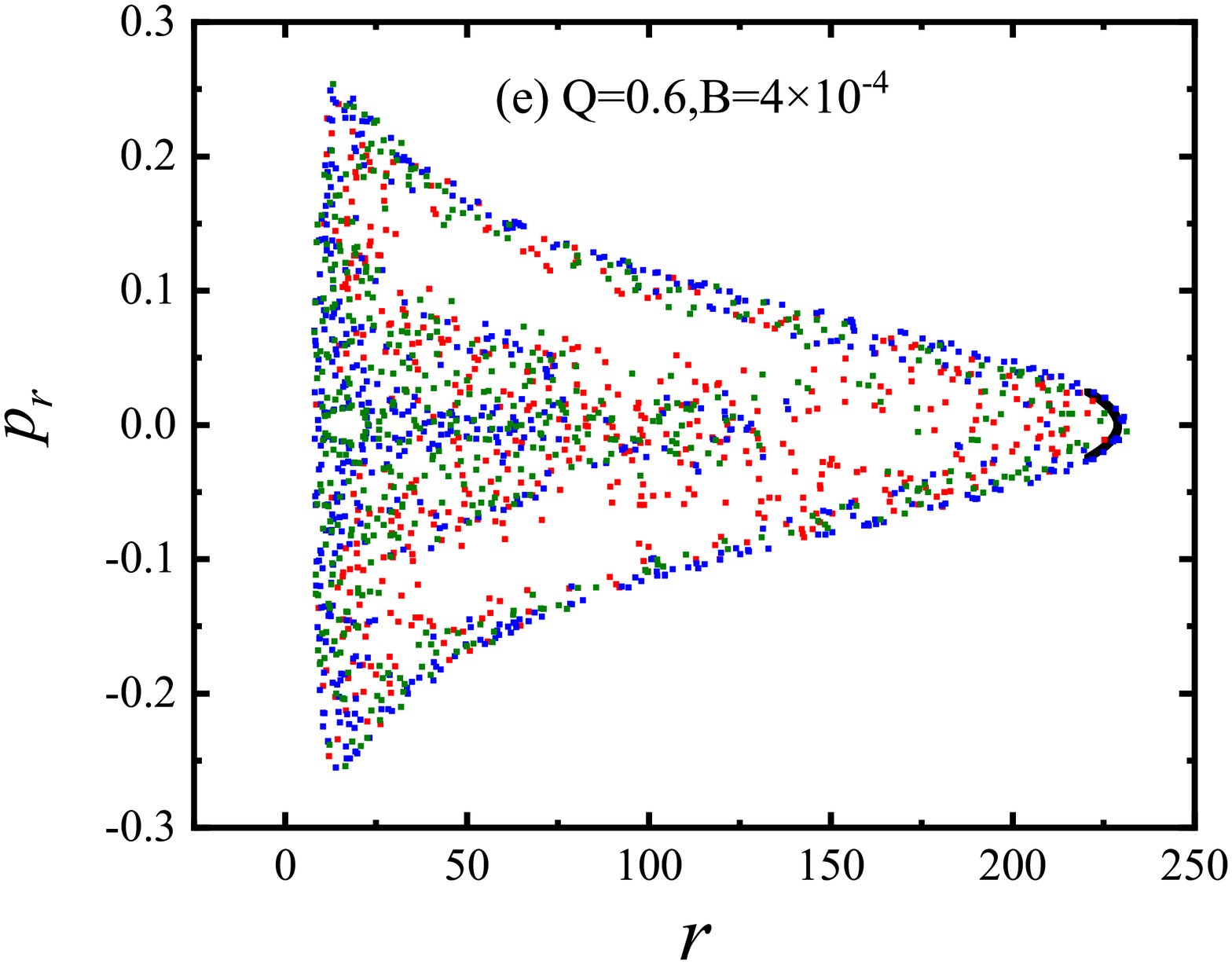}
\includegraphics[scale=0.2]{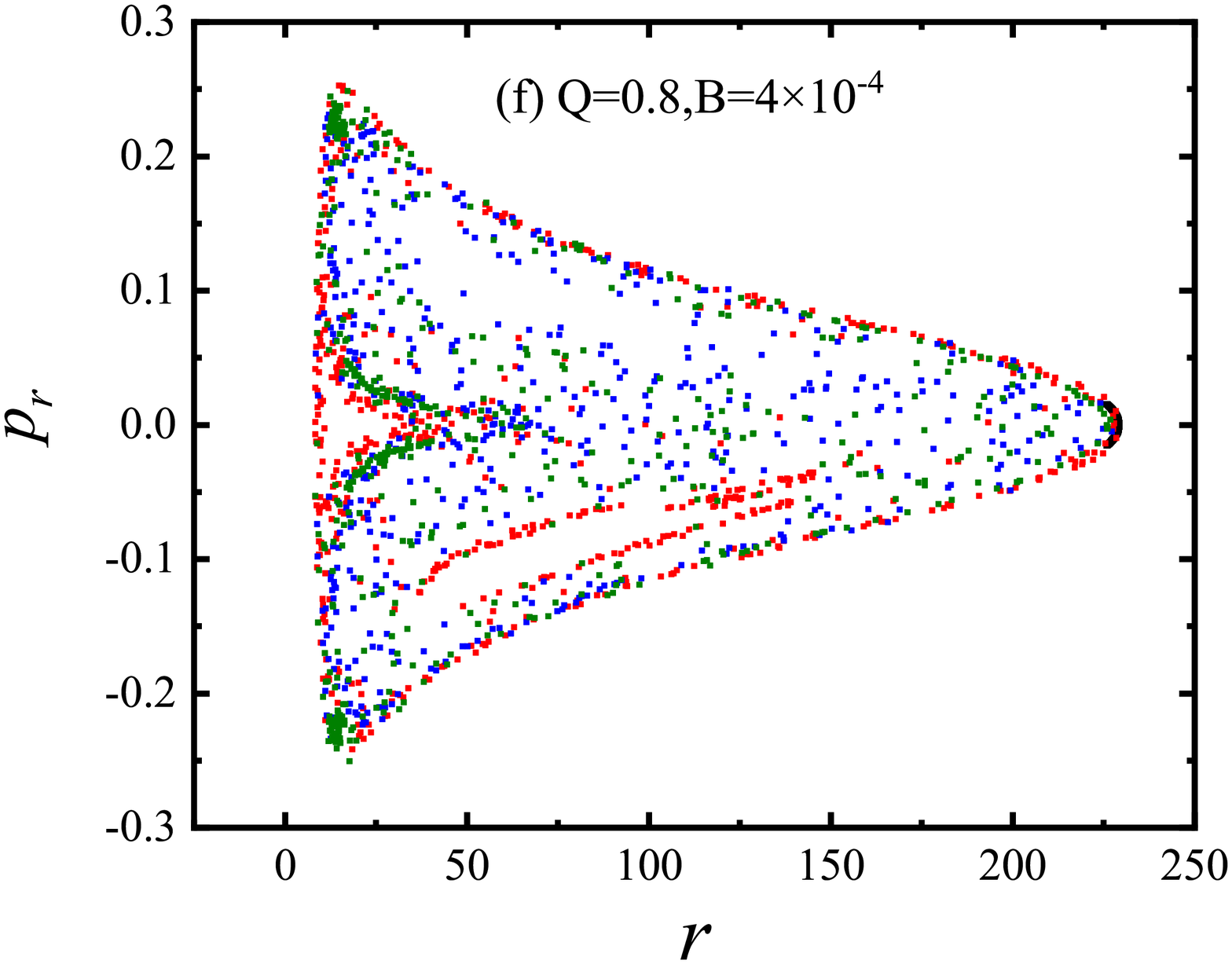}
\includegraphics[scale=0.2]{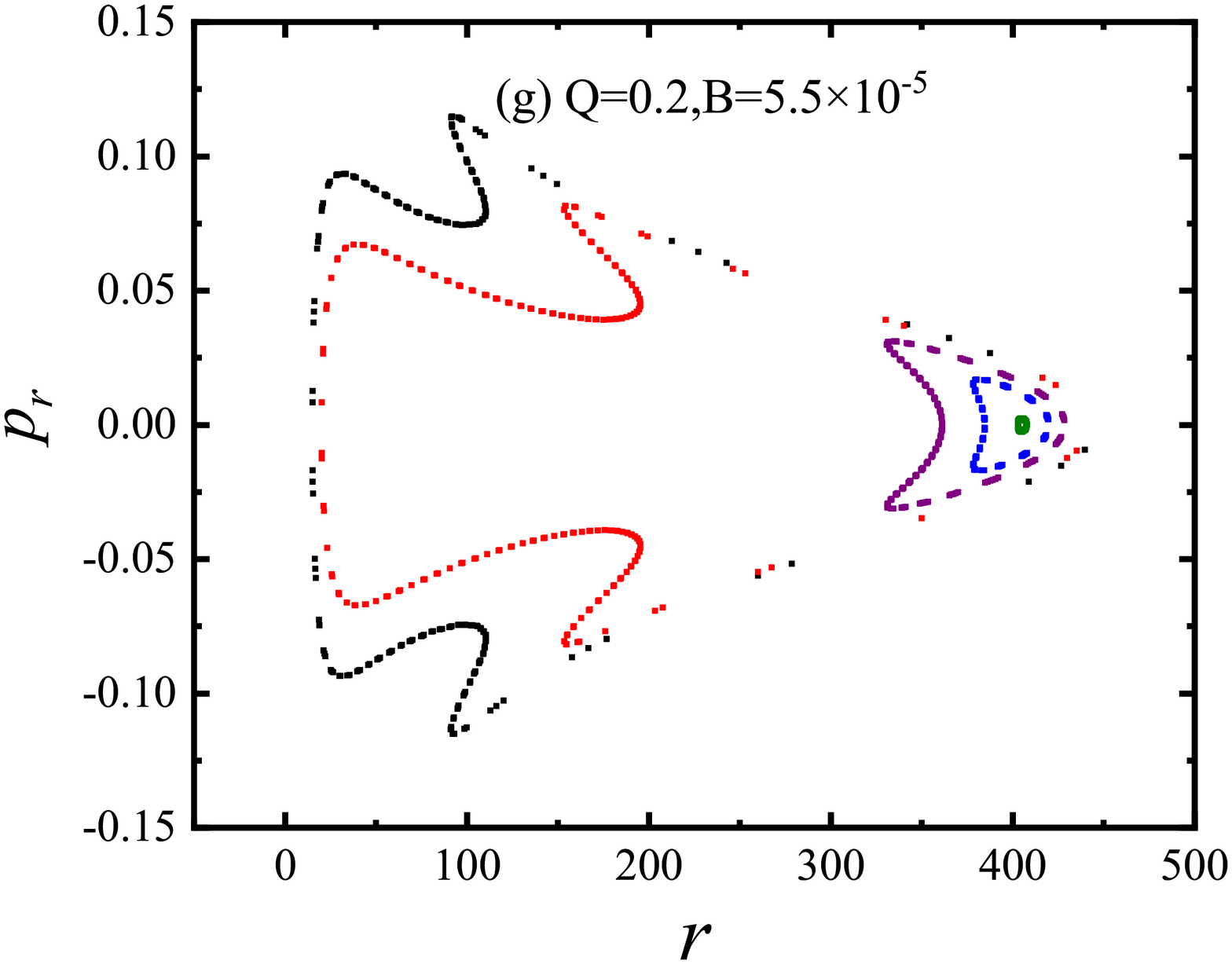}
\includegraphics[scale=0.2]{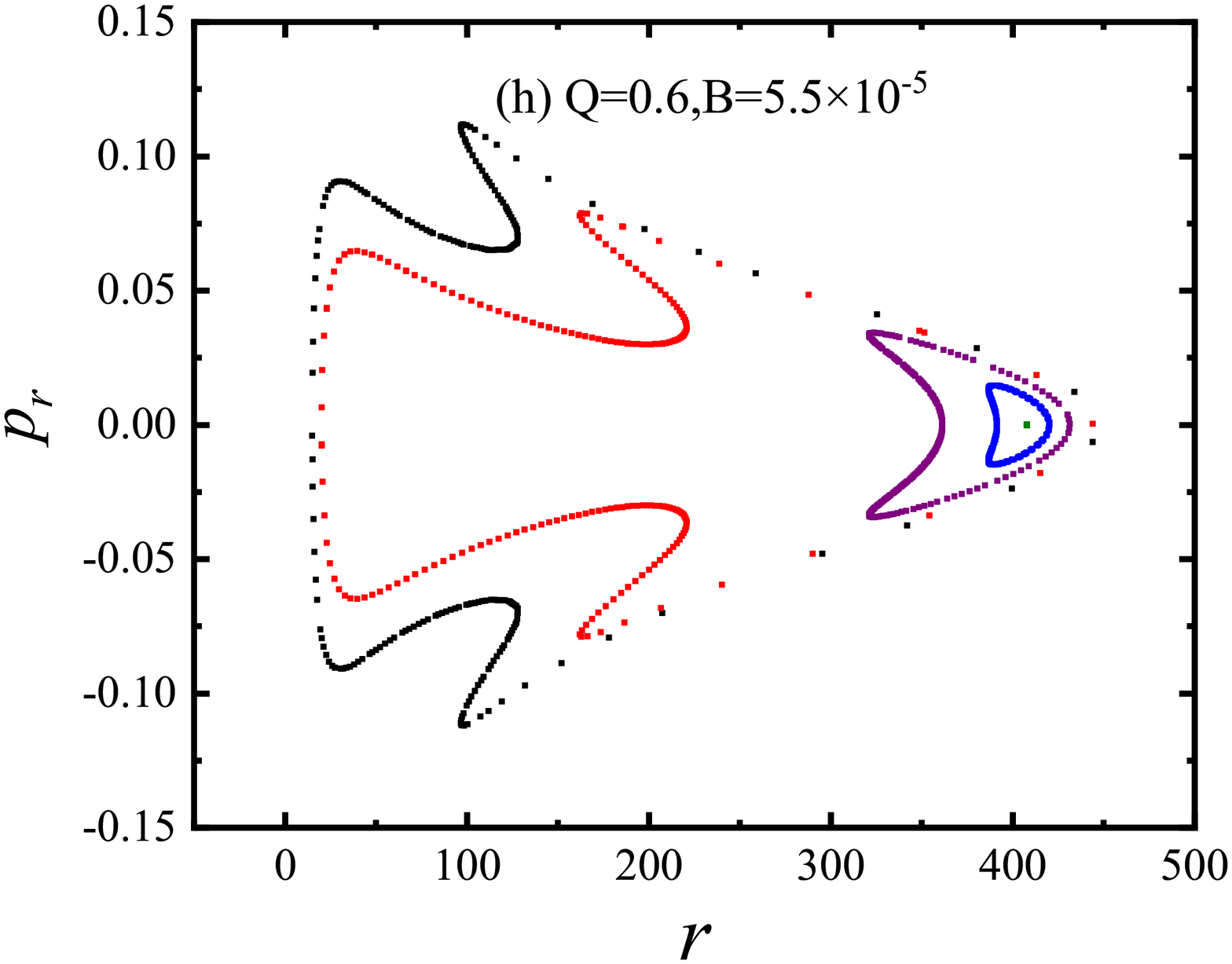}
\includegraphics[scale=0.2]{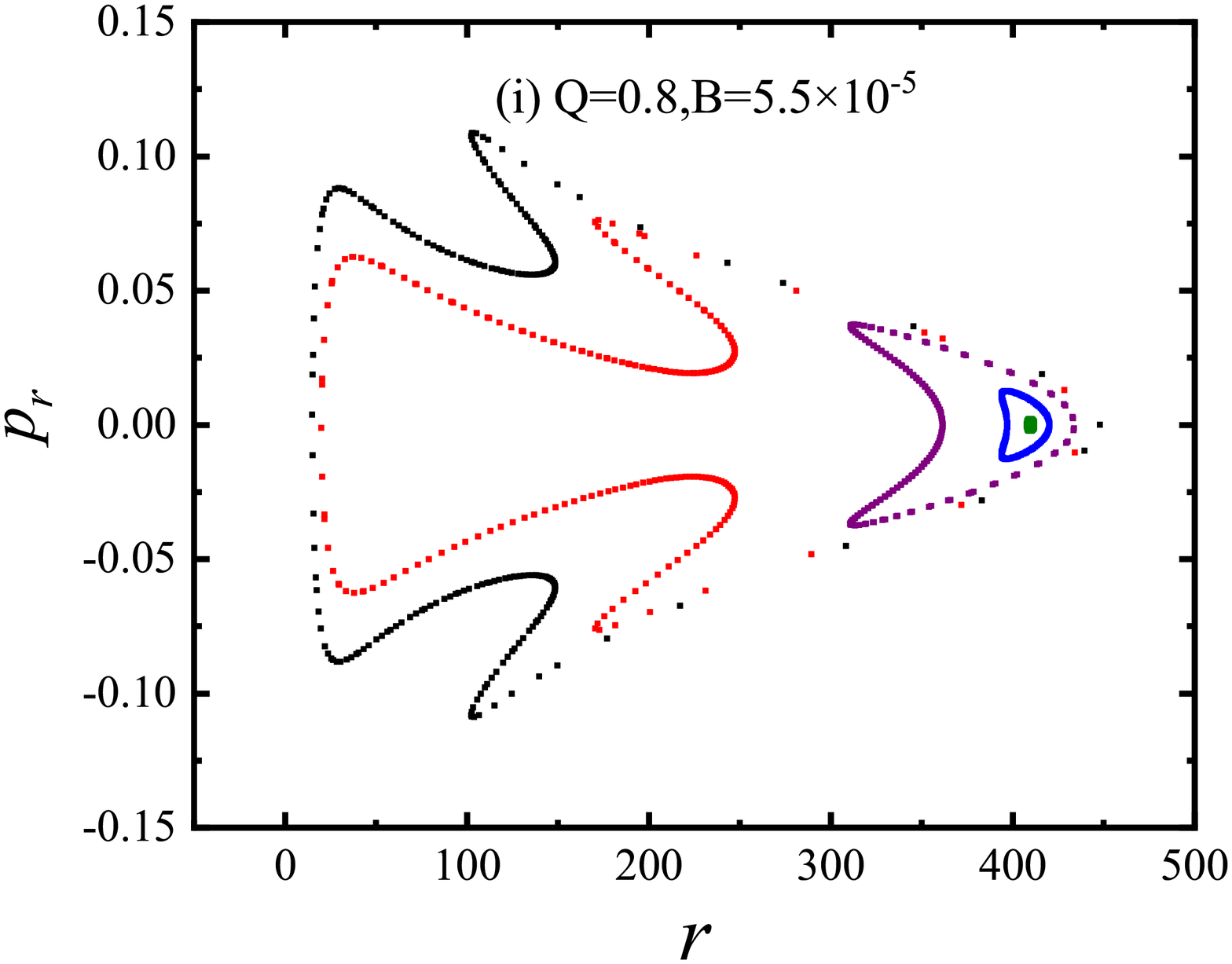}
\caption{ Poincar\'{e} sections for the motions of neutral
particles. The parameters are $E=0.998$, $L=4.5$ and $Q=0.4$ in
(a-c), but the magnetic fields are $B=5\times 10^{-5}$ in (a),
$B=3.5\times 10^{-4}$ in (b), and $B=6.5\times 10^{-4}$ in (c).
(d-i) The parameters are $E=0.998$ and $L=4.5$. Given  the
magnetic field $B=4\times 10^{-4}$, the black hole charges are
$Q=0.2$ in (d), $Q=0.6$ in (e), and $Q=0.8$ in (f). Unlike panels
(d)-(f), panels (g)-(i) replace  the magnetic field with
$B=5.5\times 10^{-5}$. The method of Poincar\'{e} sections shows
the effect of one of the varying parameters $B$ and $Q$ on the
dynamical transition, as the method of FLIs does in Fig. 4.
    }
 \label{Fig5}}
\end{figure*}

\begin{figure*}[htbp]
\center{
    \includegraphics[scale=0.2]{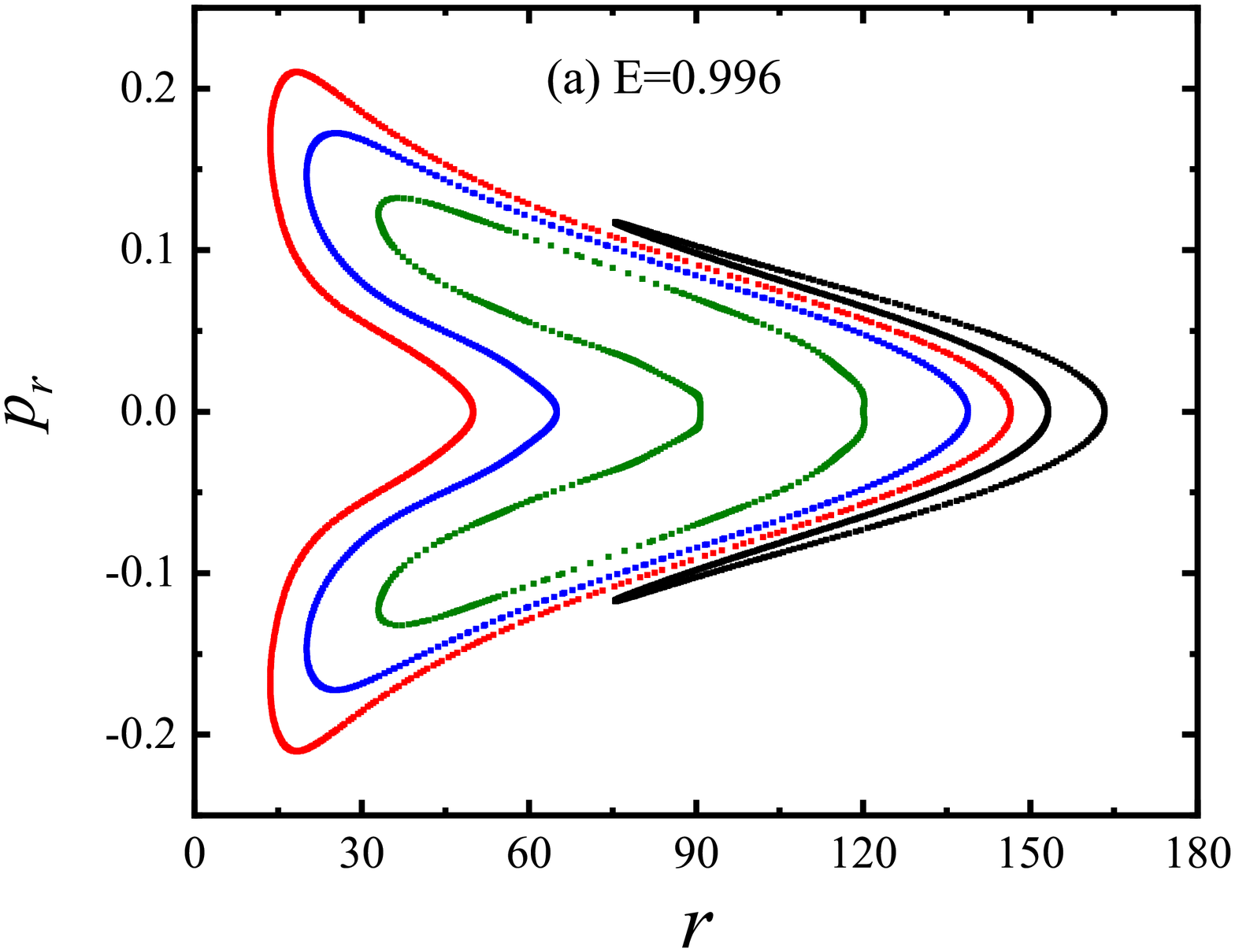}
    \includegraphics[scale=0.2]{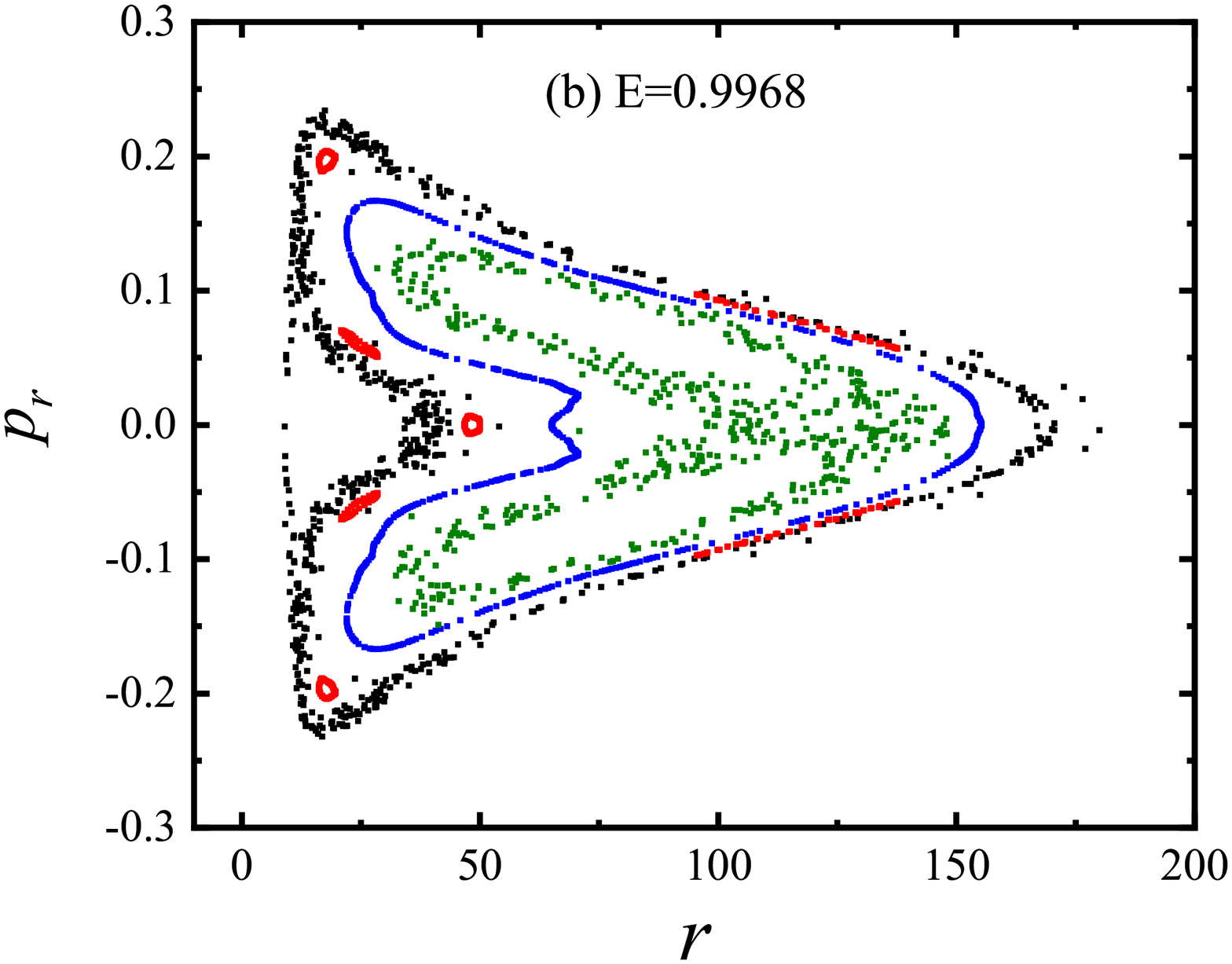}
    \includegraphics[scale=0.2]{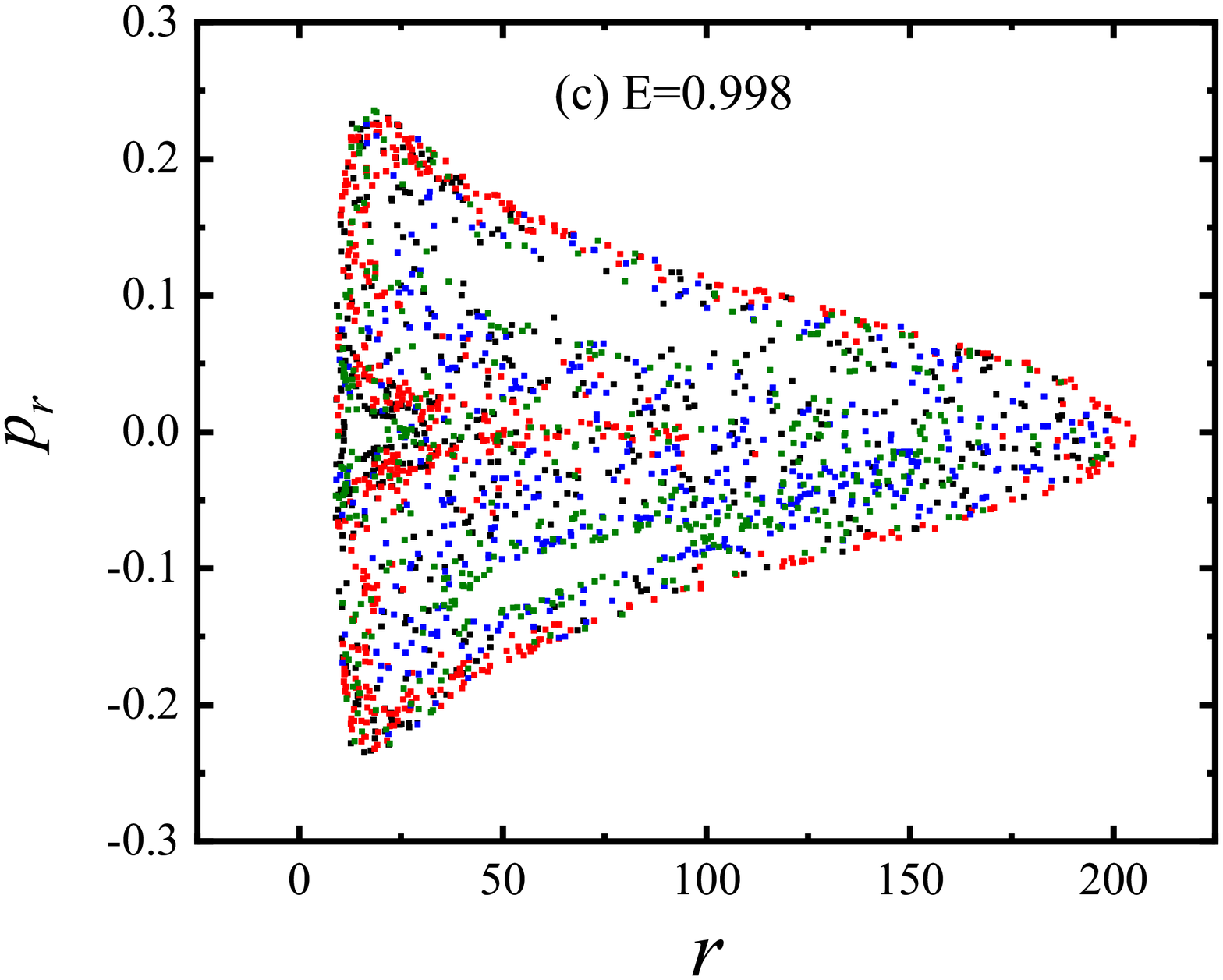}
    \includegraphics[scale=0.2]{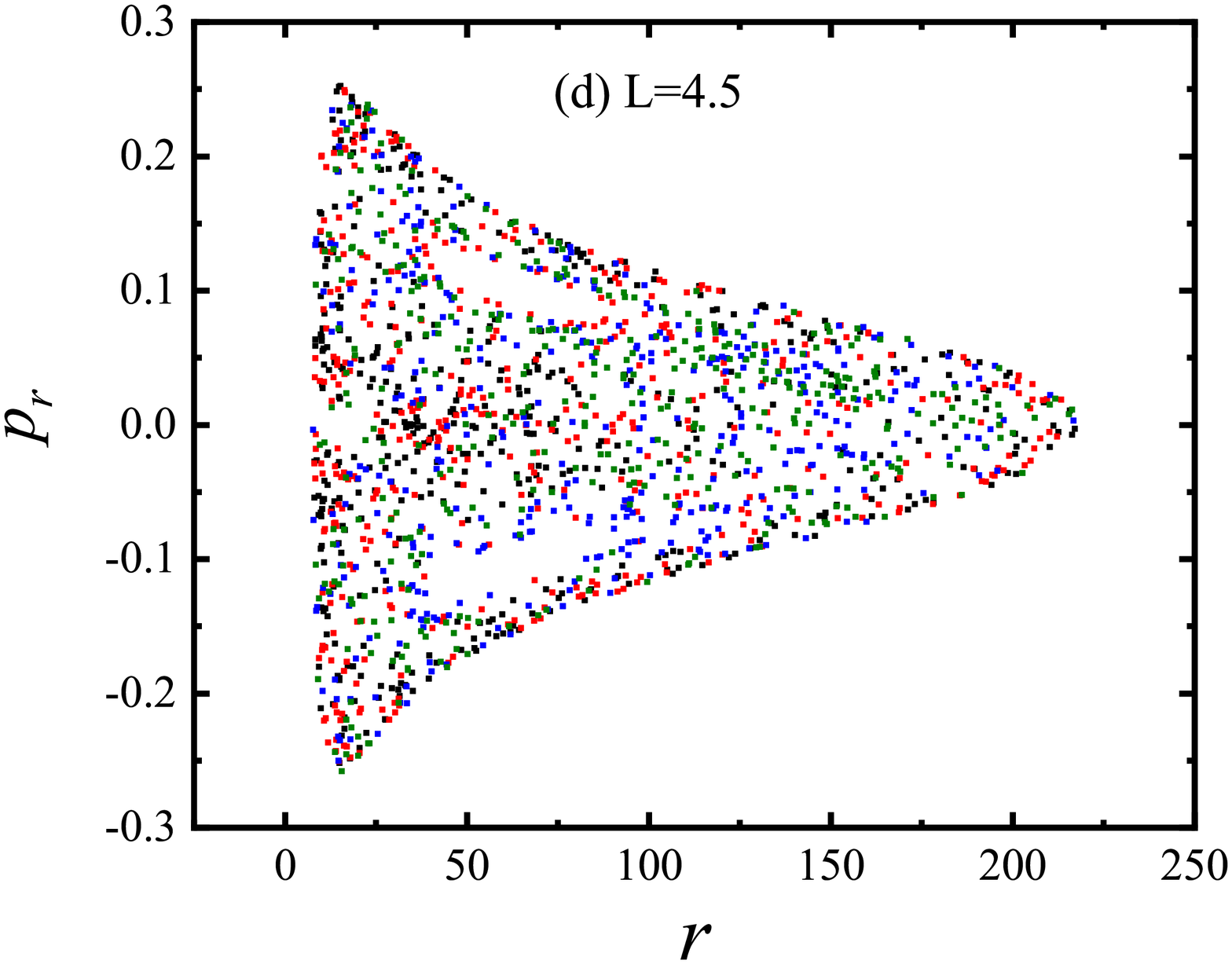}
    \includegraphics[scale=0.2]{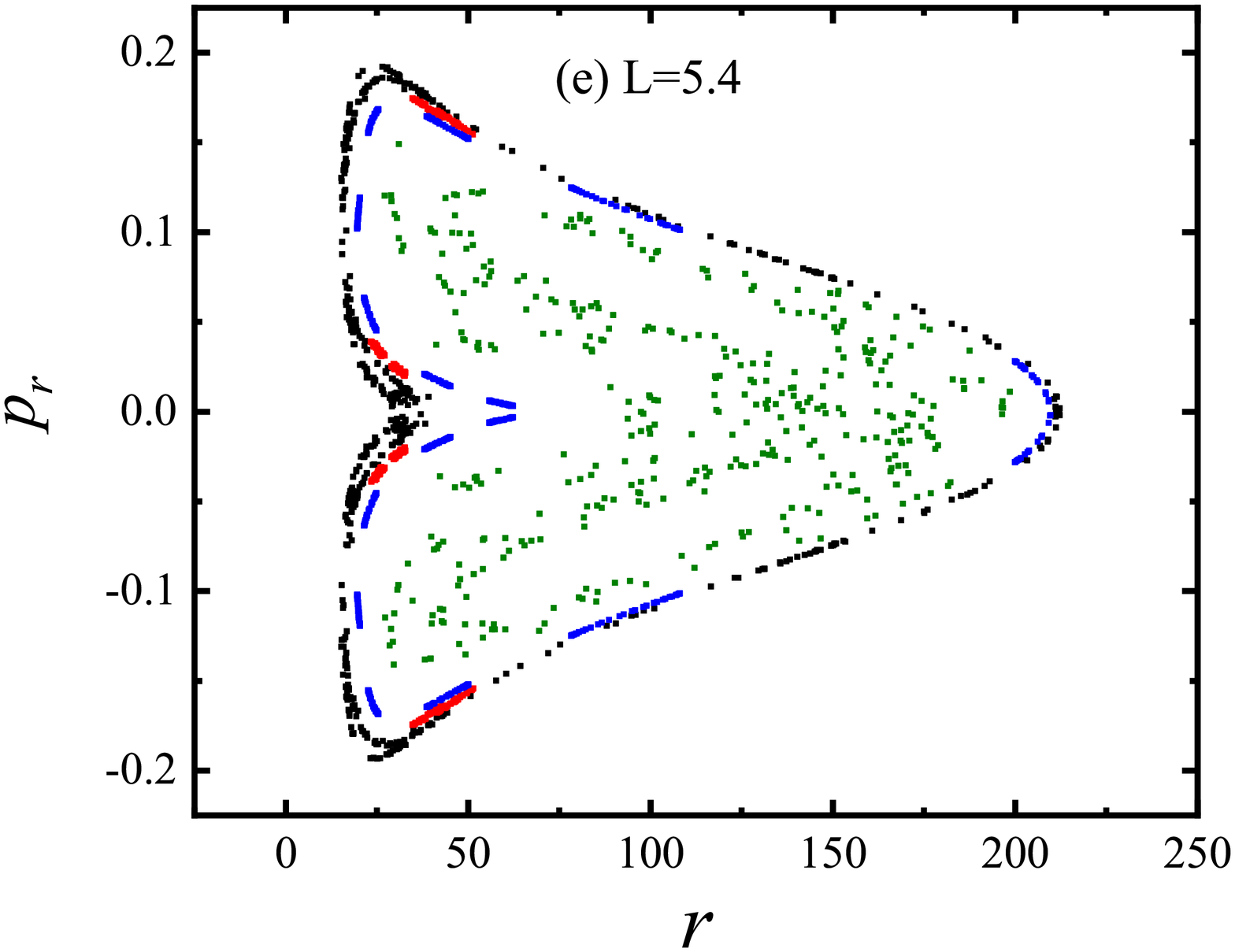}
    \includegraphics[scale=0.2]{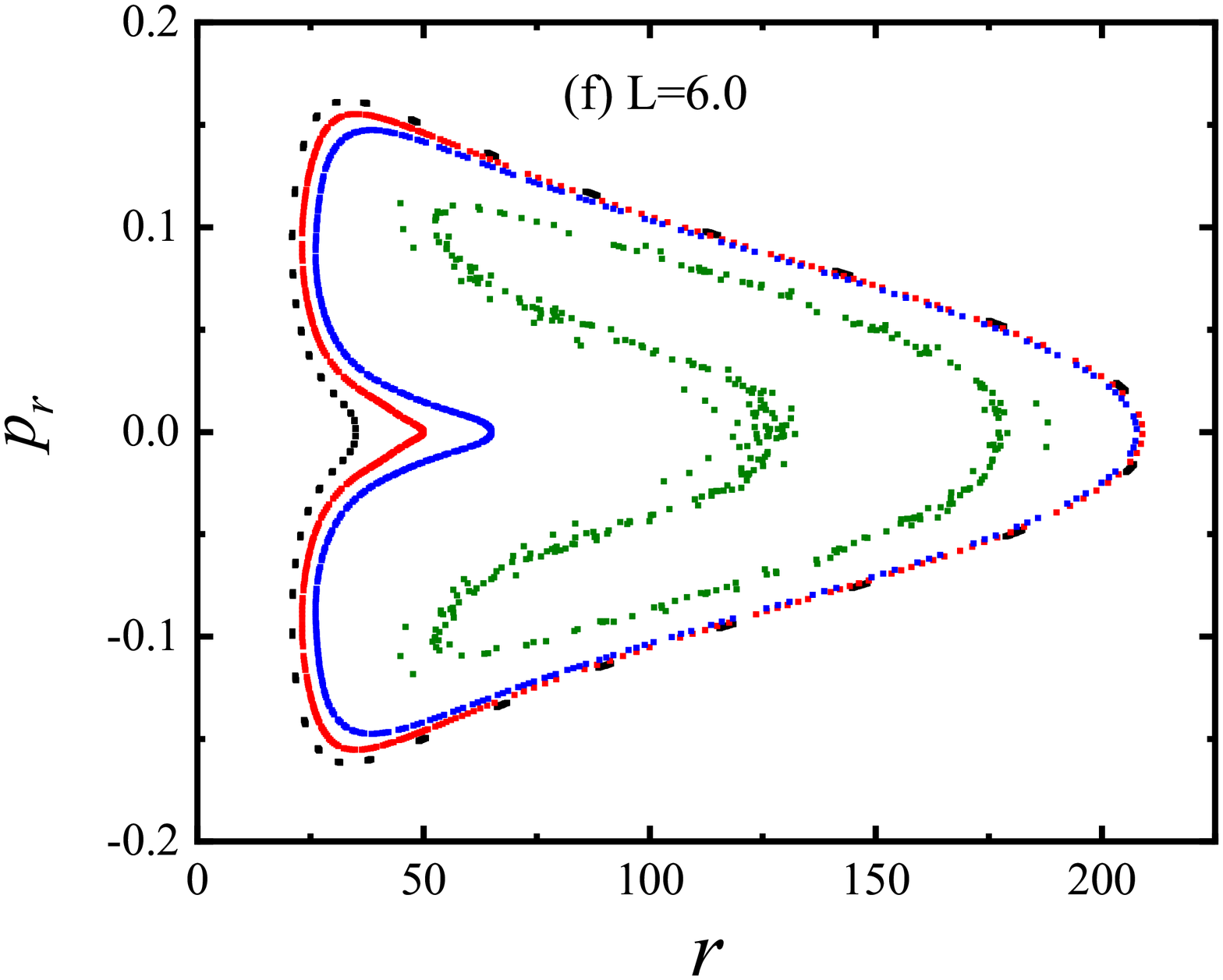}
    \caption{Fig. 5 continued.
(a-c) The parameter values are $B=5 \times10^{-4} $, $Q=0.4$ and
$L=4.7$, but the energy values are (a) $E=0.996$, (b) $E=0.9968$
and (c) $E=0.998$. (d-f) The parameter values are $E=0.998 $,
$Q=0.4$, and $B=4.5 \times10^{-4}$, while the  angular momentum
values are (d) $L=4.5$, (e) $L=5.5$, and (f) $L=6$. It is clear
that the degree of chaos increases with the energy increasing,
whereas decreases with the angular momentum increasing.
      }
 \label{Fig6}}
\end{figure*}

\begin{figure*}[htbp]
    \center{
    \includegraphics[scale=0.2]{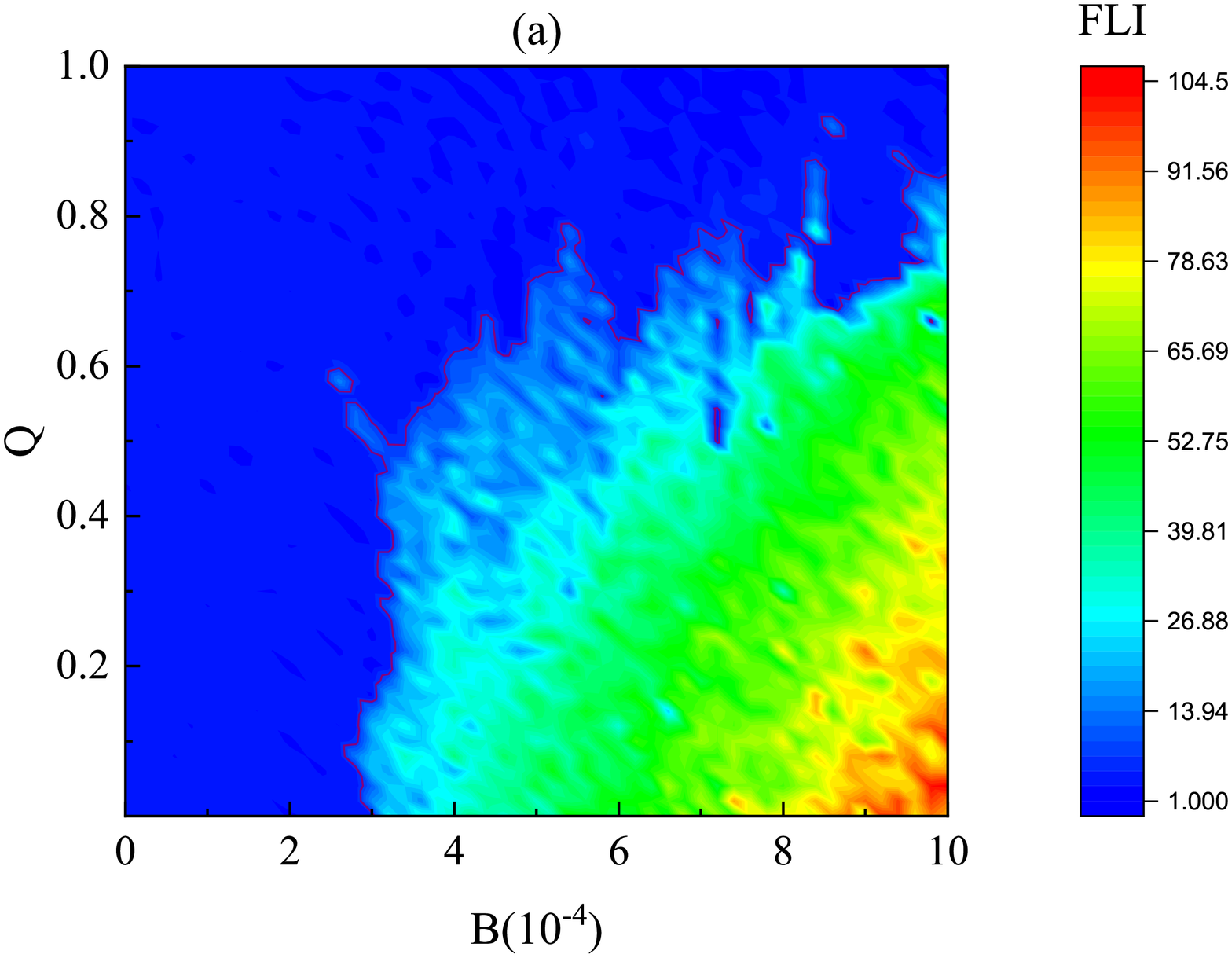}
    \includegraphics[scale=0.2]{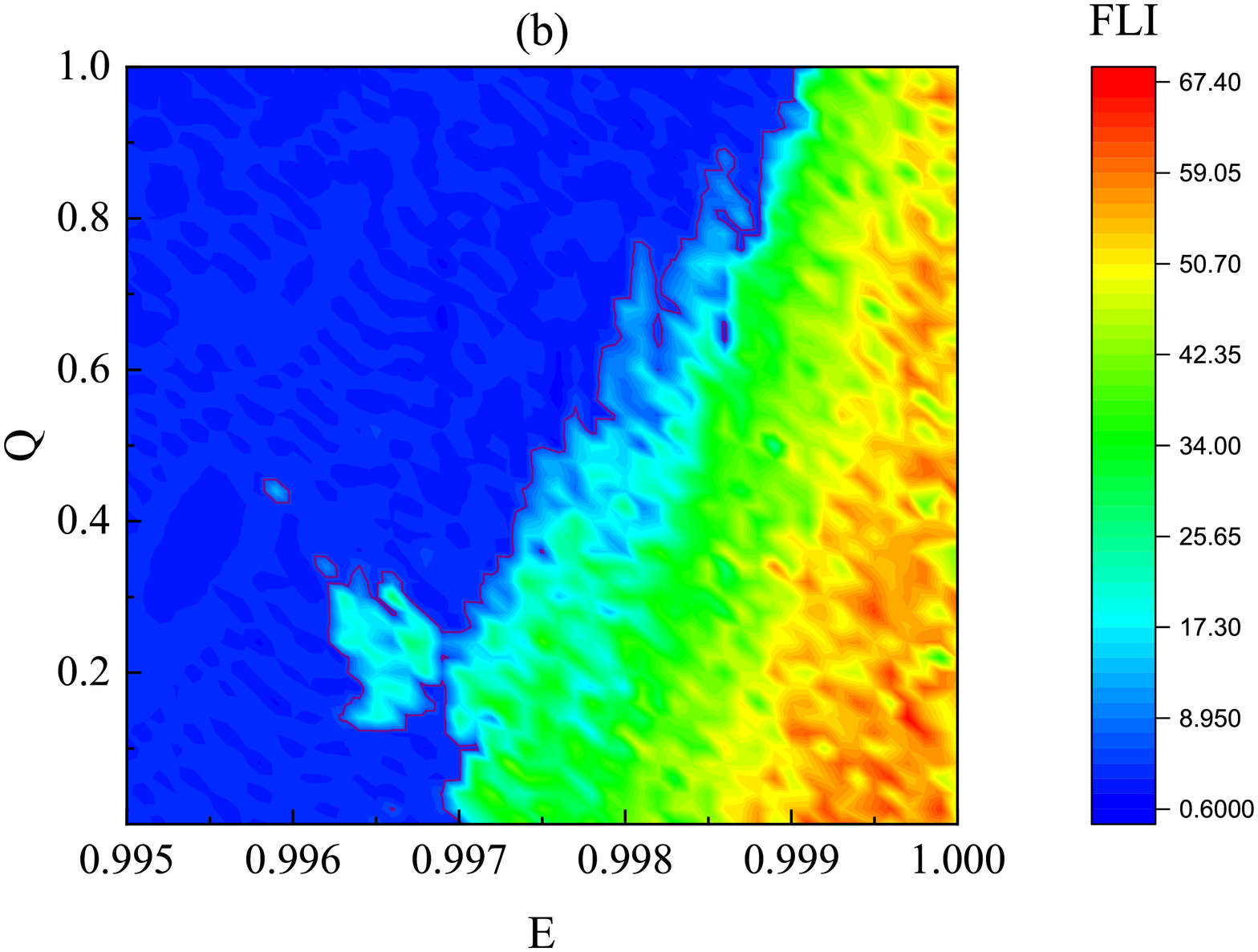}
    \includegraphics[scale=0.2]{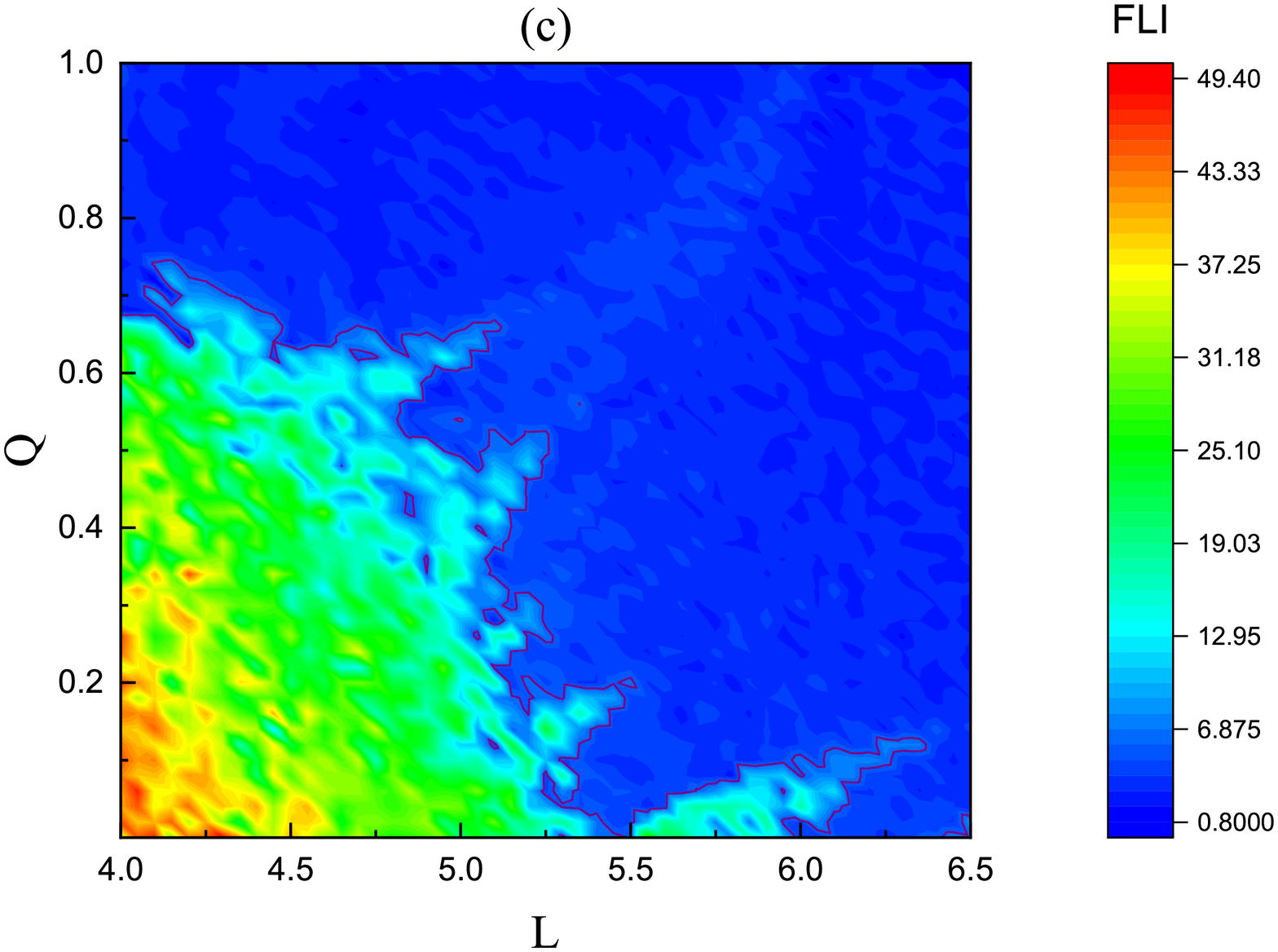}
    \caption{Same as Fig. 4, but the particle charge $q=0$ is
    replaced with $q=0.5$. Fig. 7 (a), (b) and (c) correspond to Fig. 4 (a), (b) and
    (c), respectively. Under some circumstances, chaos gets stronger as the magnetic field strength $B$ and energy
    $E$ increase, but weaker when the black hole charge $Q$ or the
    particle angular momentum $L$ increases. }
     \label{Fig7}}
\end{figure*}

\begin{figure*}[htbp]
    \center{
        \includegraphics[scale=0.2]{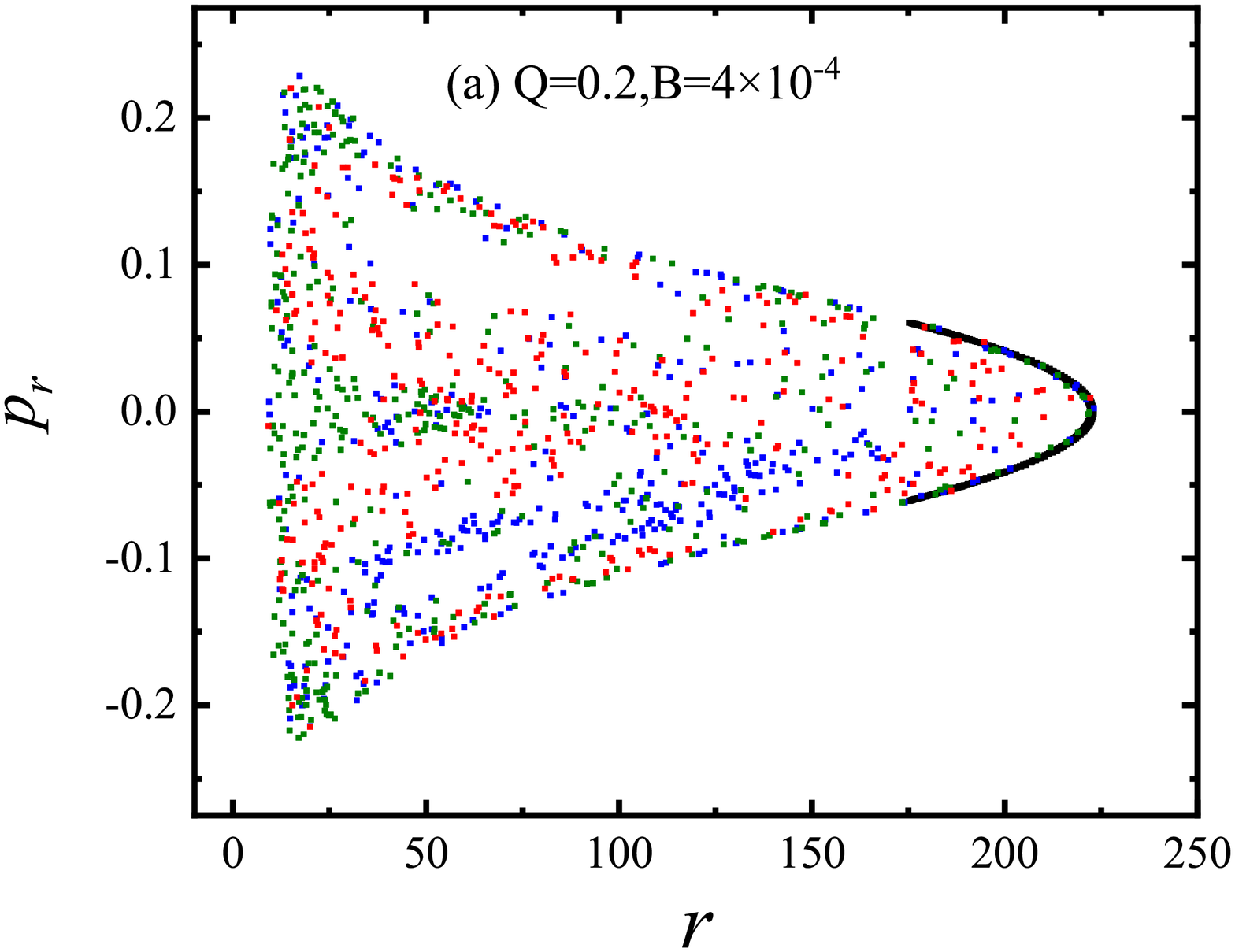}
        \includegraphics[scale=0.2]{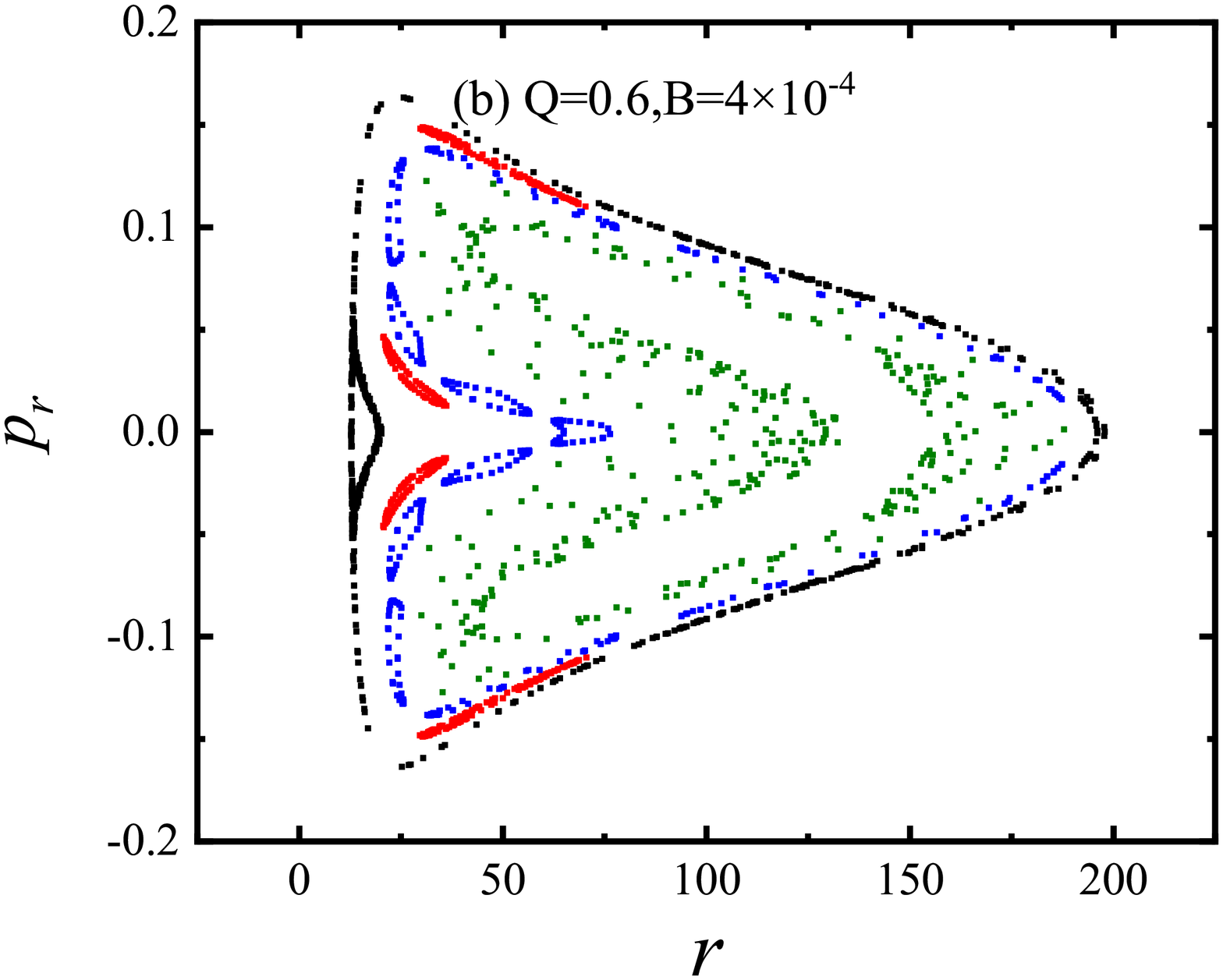}
        \includegraphics[scale=0.2]{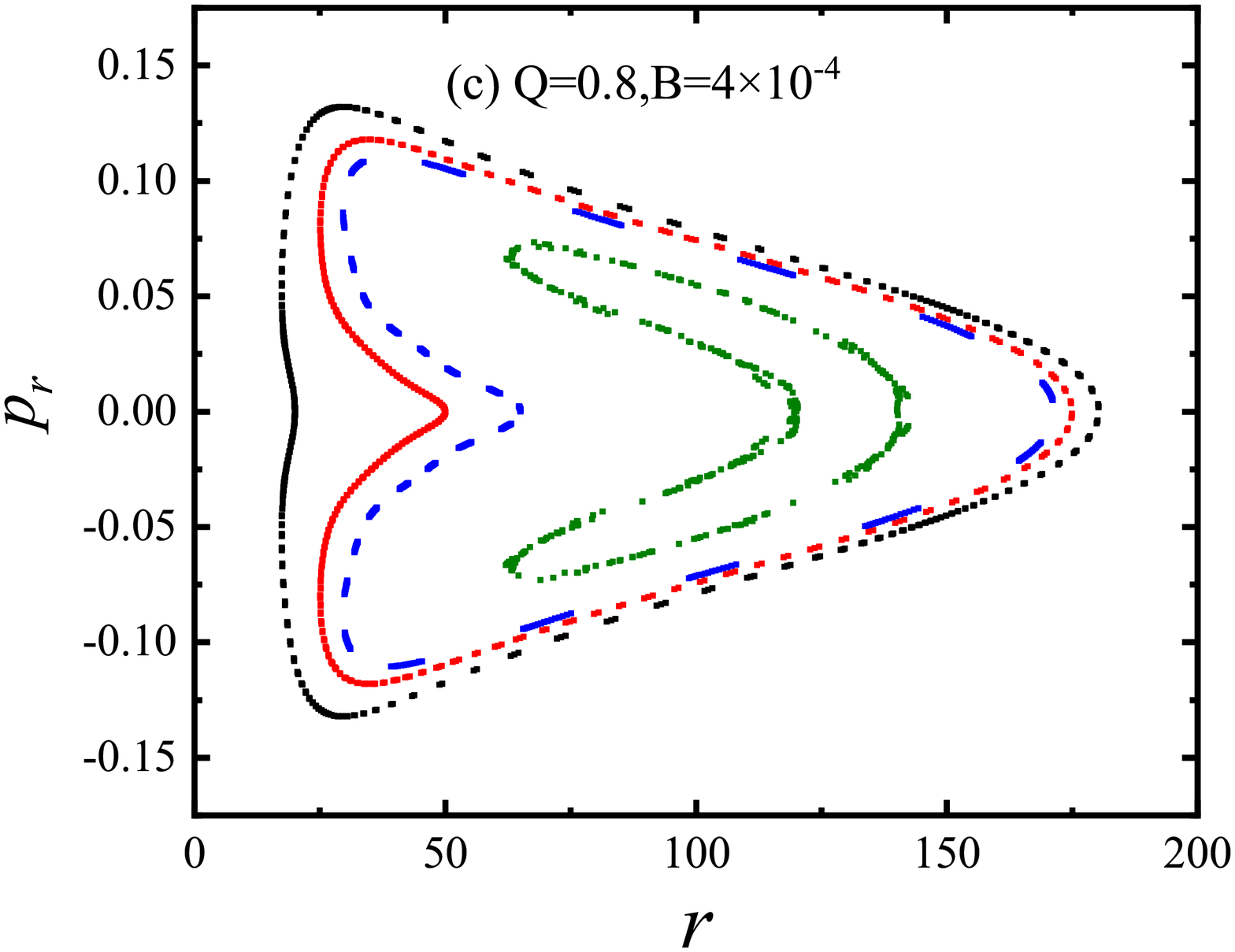}
        \includegraphics[scale=0.2]{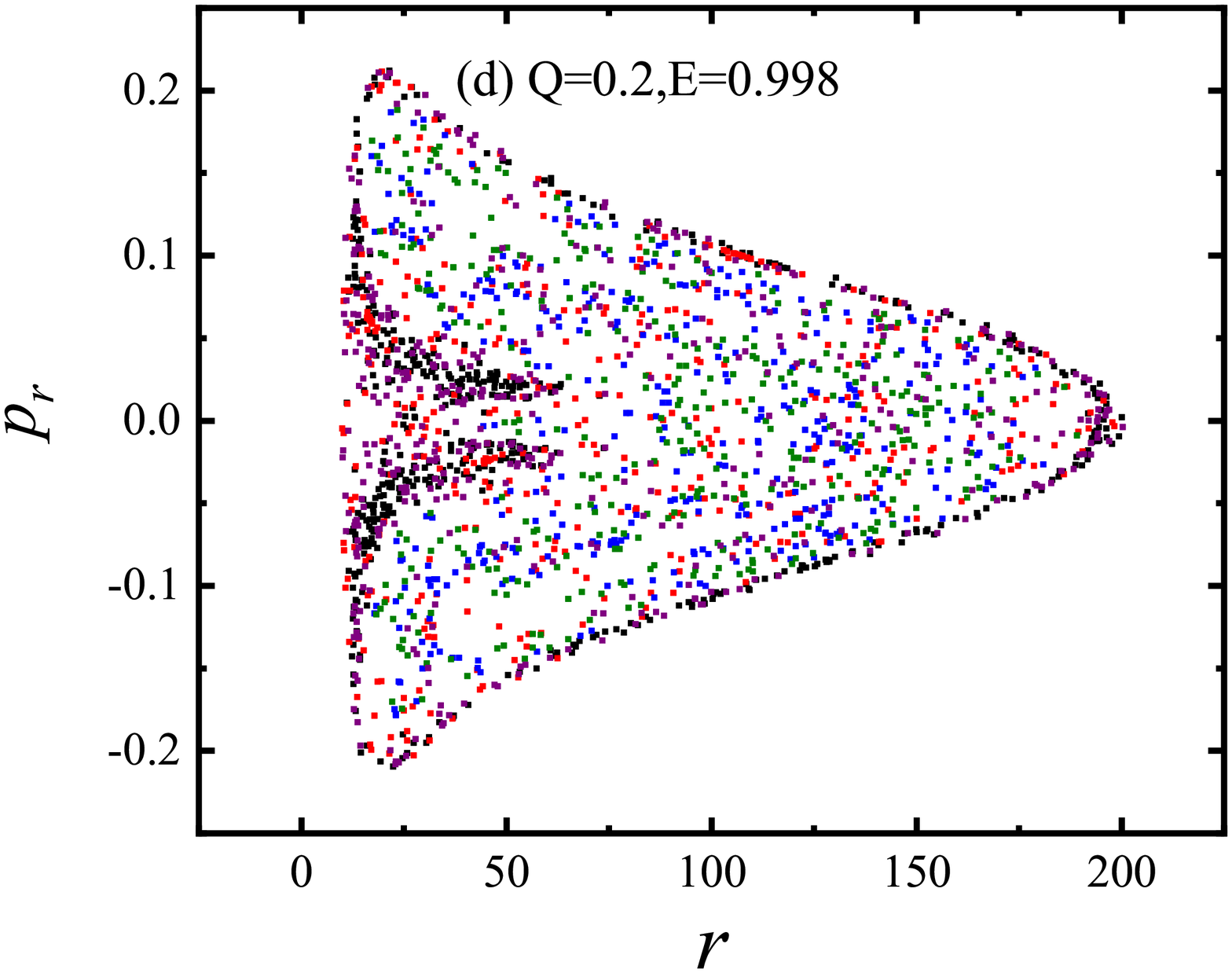}
        \includegraphics[scale=0.2]{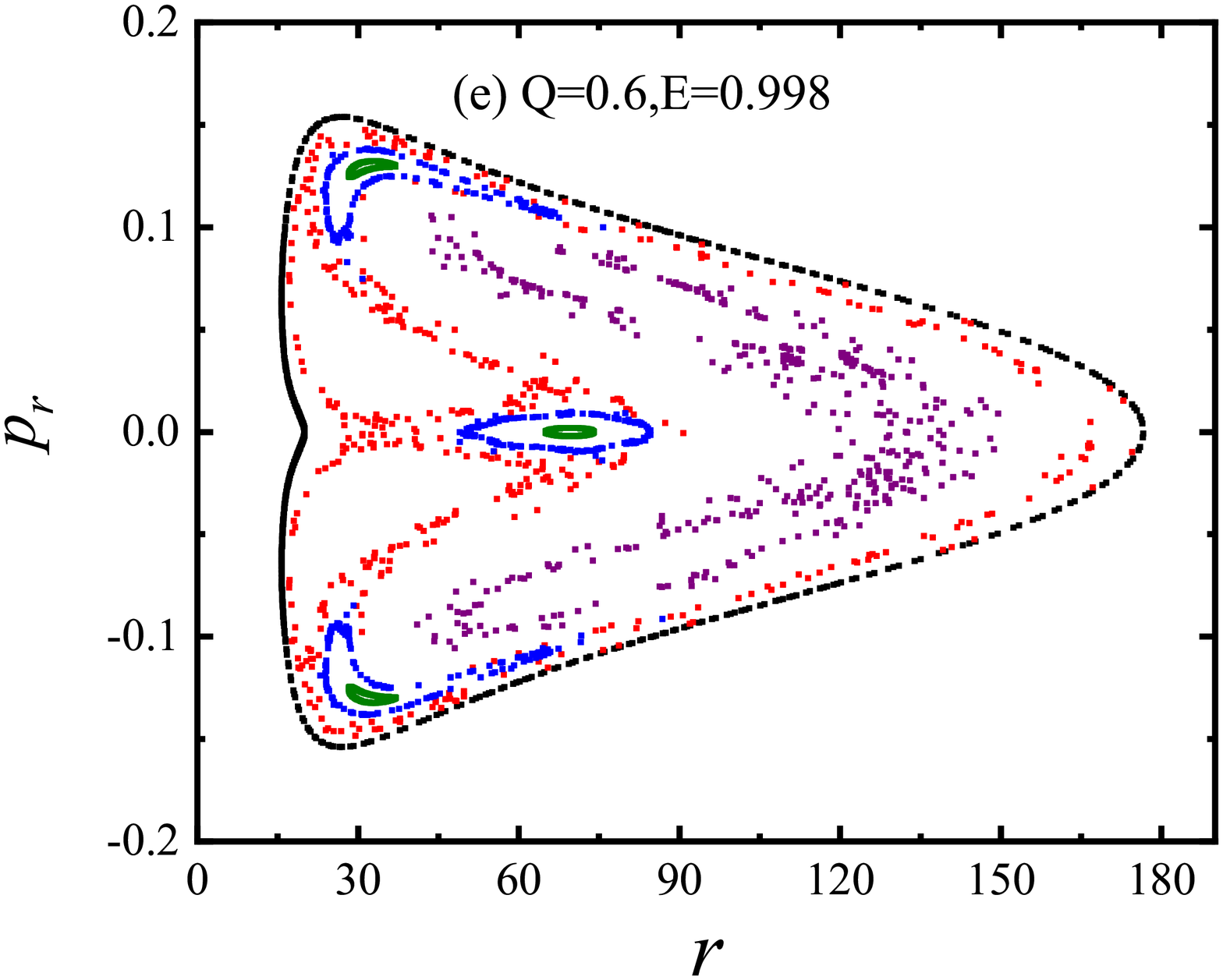}
        \includegraphics[scale=0.2]{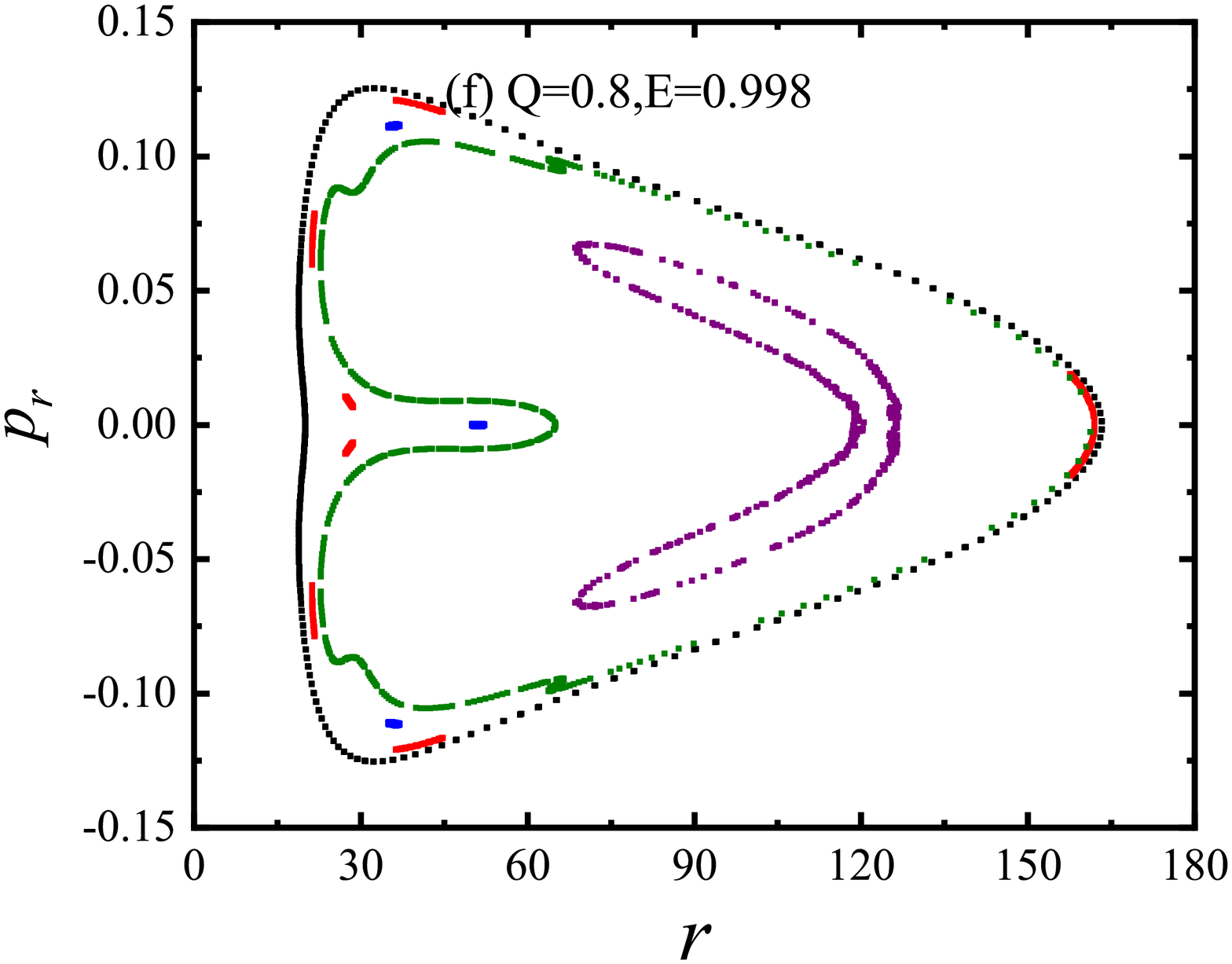}
        \includegraphics[scale=0.2]{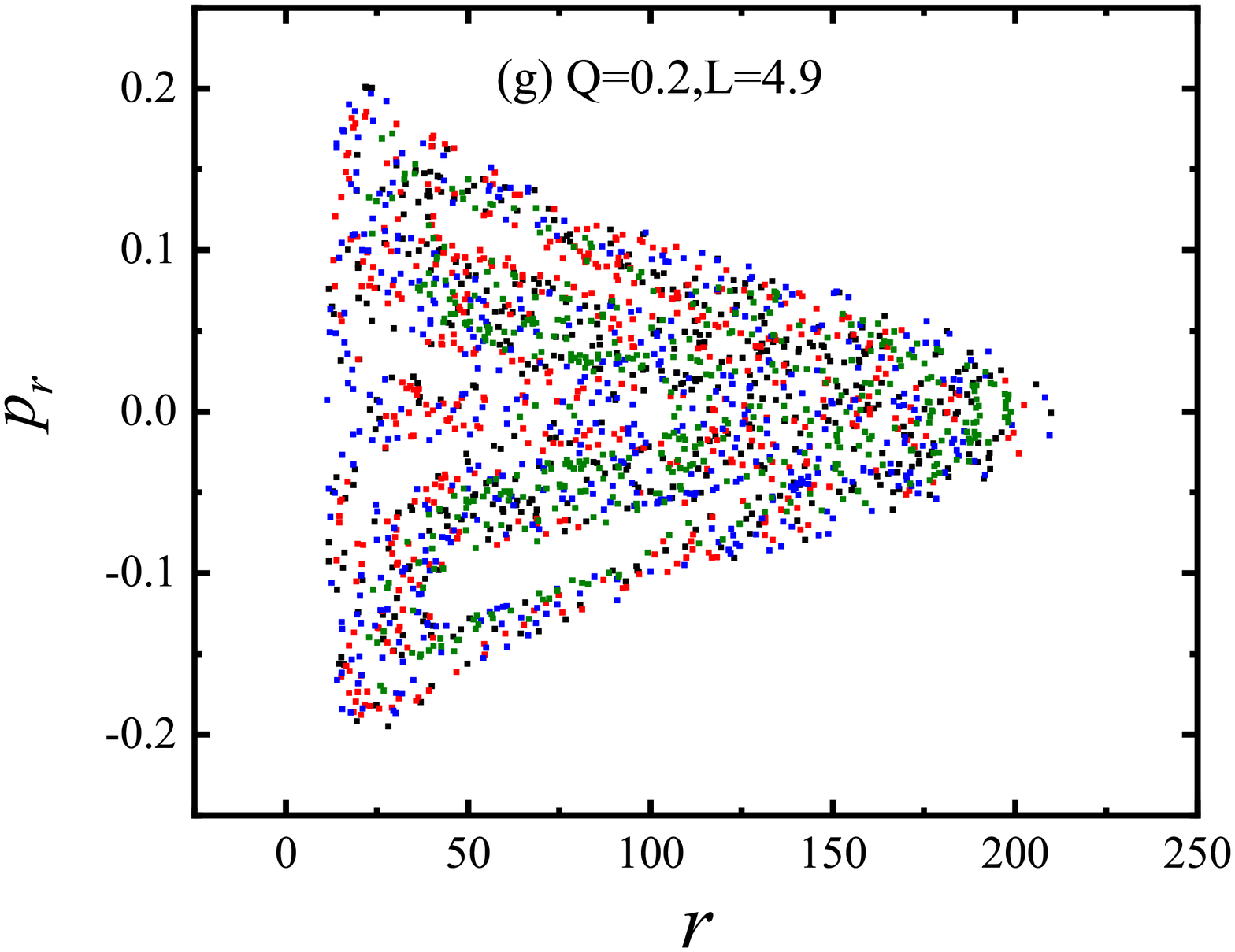}
        \includegraphics[scale=0.2]{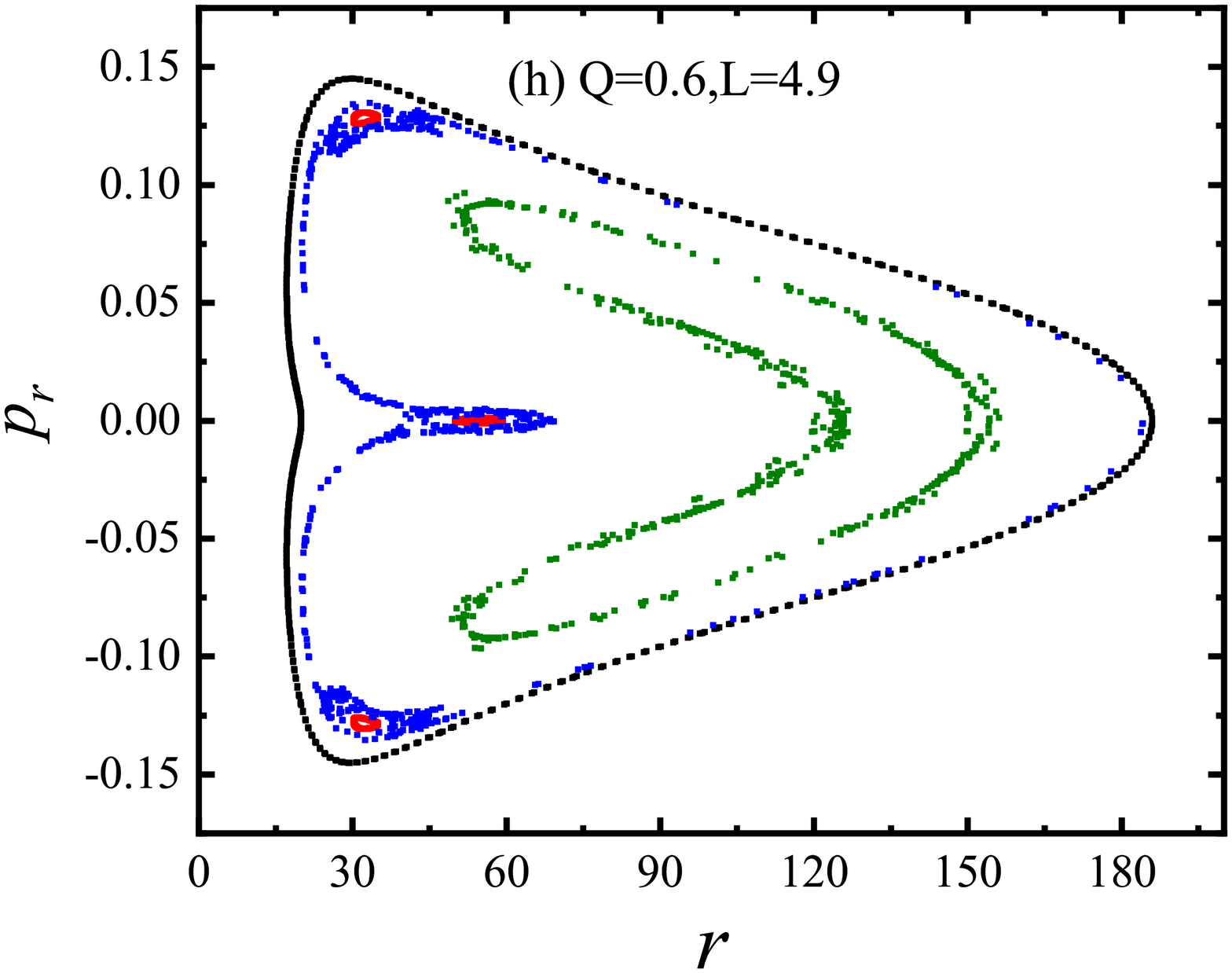}
        \includegraphics[scale=0.2]{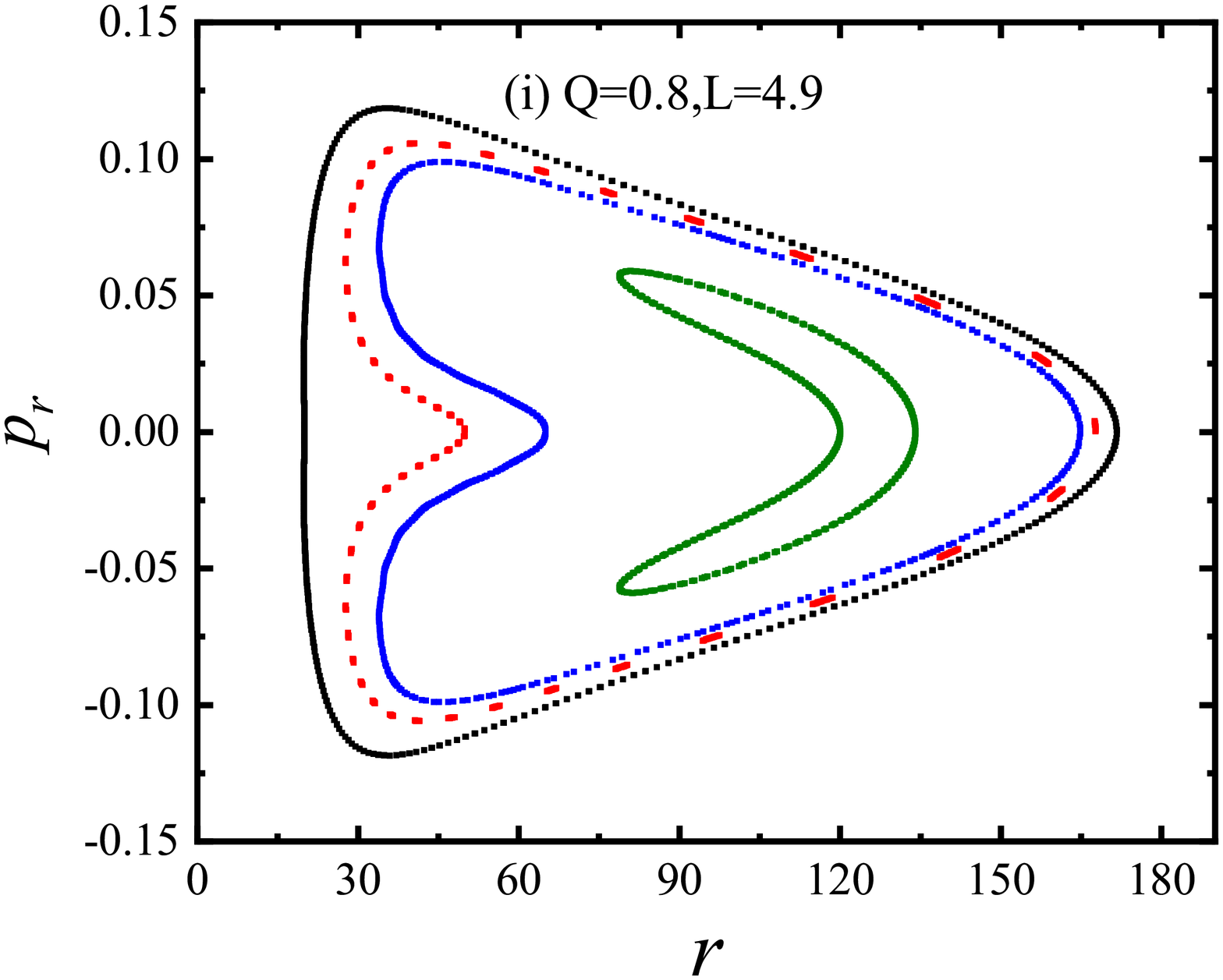}
        \caption{
Poincar\'{e} sections. The parameters are the particle energy
$E=0.998$ and the particle charge $q=0.5$. The black hole charges
are  $Q=0.2$ in panels (a), (d) and (g), $Q=0.6$ in panels (b),
(e) and (h), and $Q=0.8$ in panels (c), (f) and (i). (a-c) The
magnetic field strength is $B=4\times10^{-4}$, and the other
parameters are those of Fig. 7(a). (d-f) The energy is $E=0.998$,
and the other parameters are those of Fig. 7(b). (g-i) The
particle angular momentum is $L=4.9$, and the other parameters are
those of Fig. 7(c). For the three cases, chaos becomes weaker as
the black hole charge increases.}
     \label{Fig8}}
\end{figure*}

\begin{figure*}[htbp]
    \center{
        \includegraphics[scale=0.3]{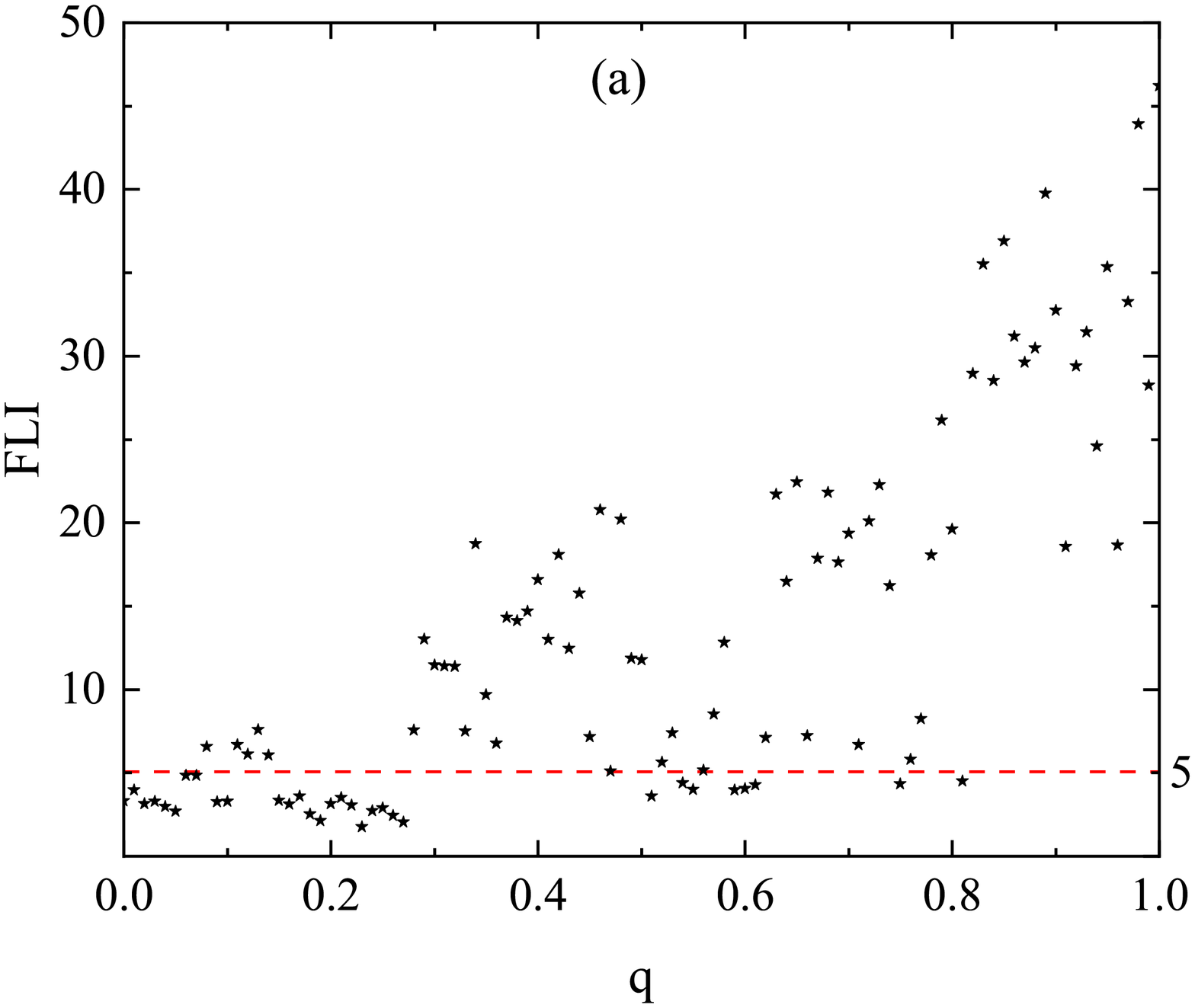}
        \includegraphics[scale=0.3]{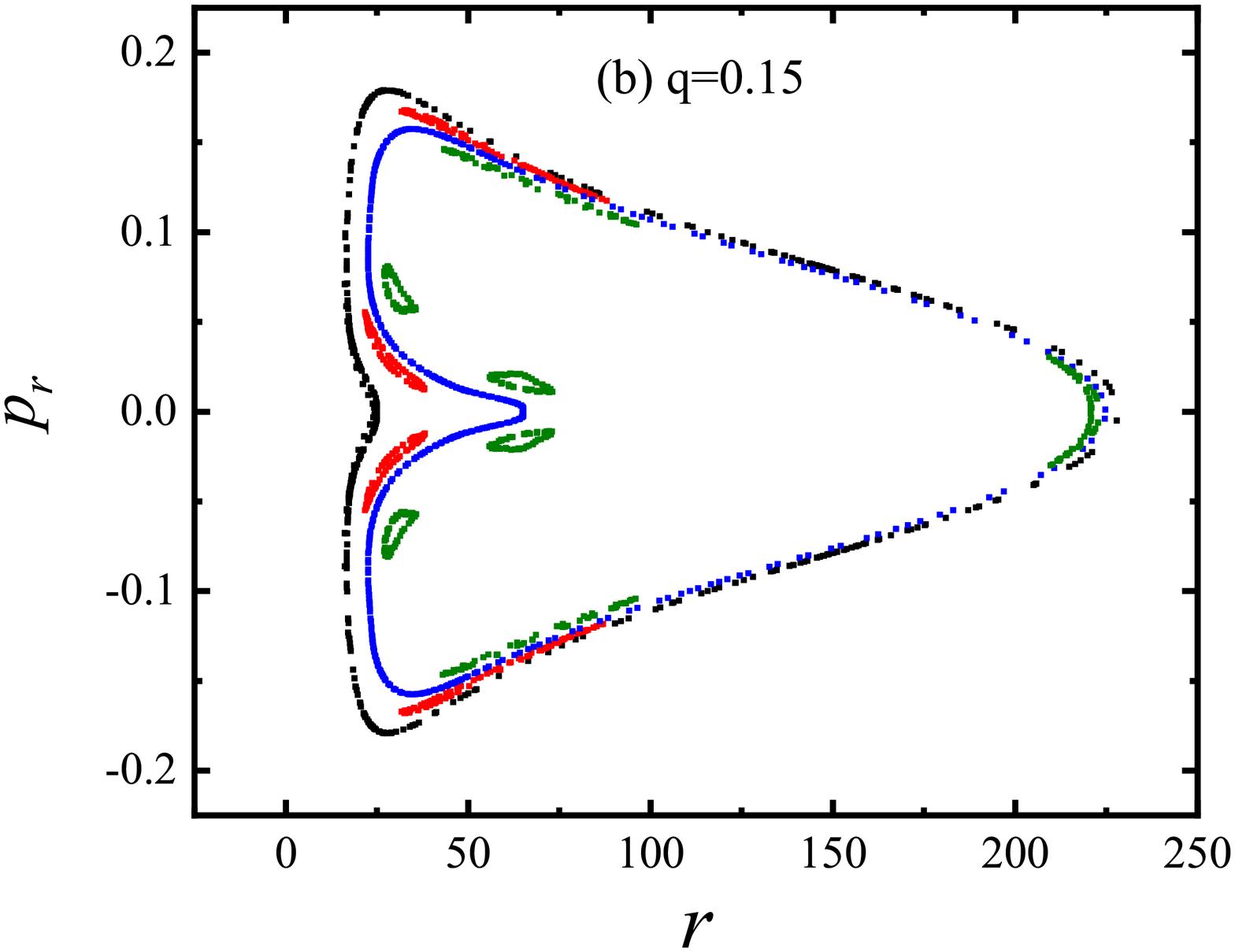}
        \includegraphics[scale=0.3]{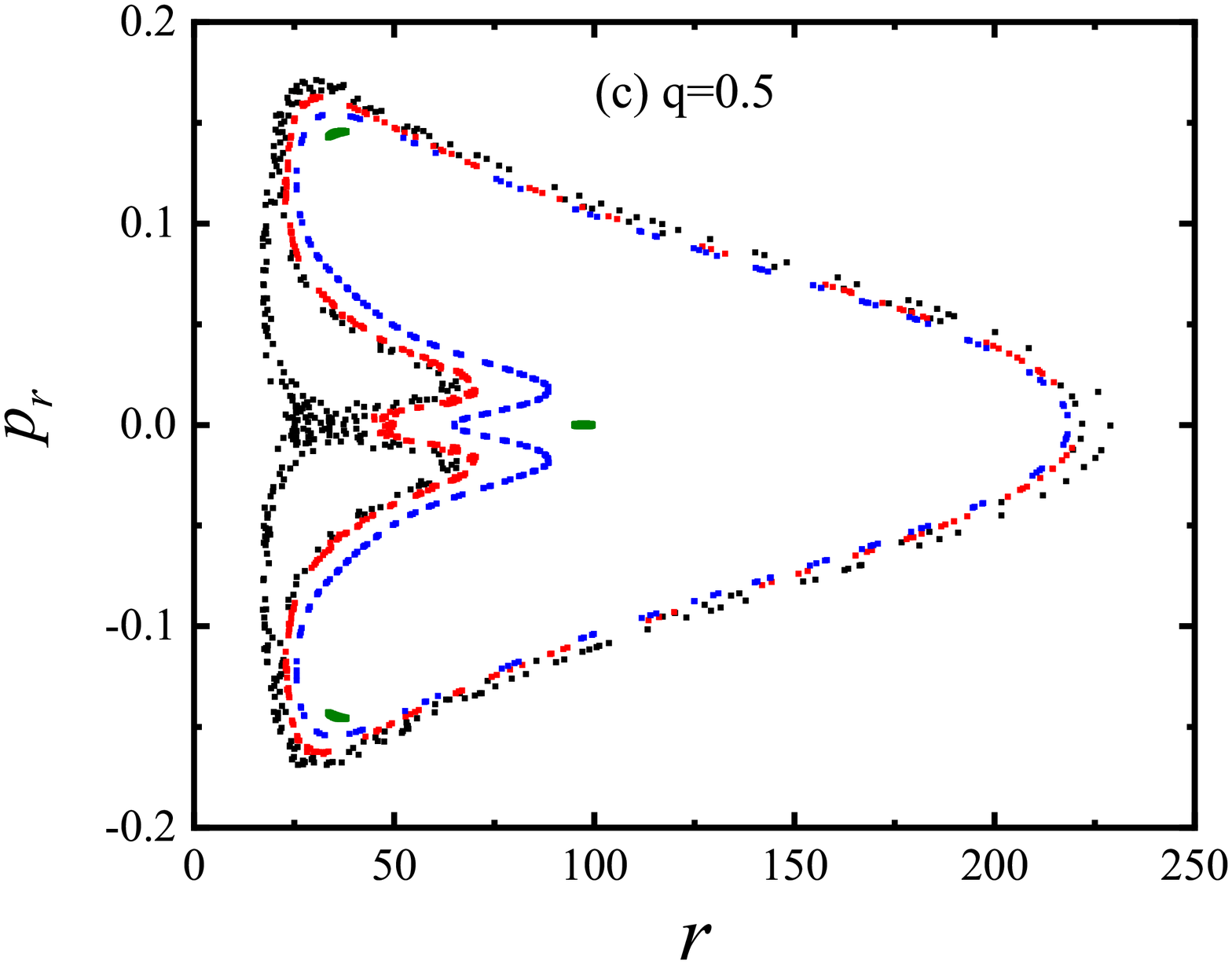}
        \includegraphics[scale=0.3]{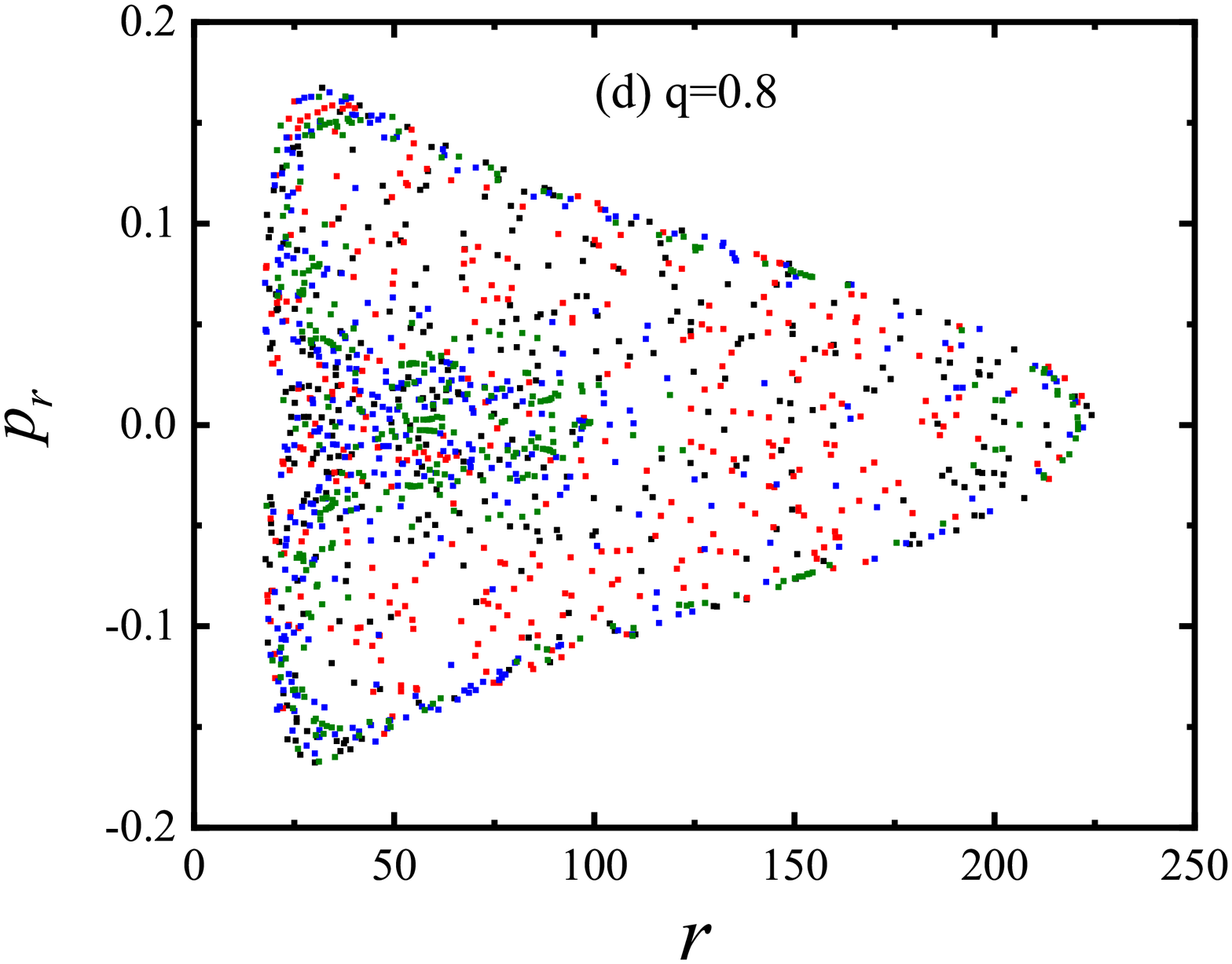}
        \caption{
(a) The dependence of FLI on the particle positive charge $q$. The
initial separation is $r=40$, and the parameters are $E=0.998$,
$L=5.7$, $B=4\times 10^{-4}$, and $Q=0.1$. (b-d) Poincar\'{e}
sections. The particle charges are (a) $q=0.15$, (b) $q=0.5$ and
(c) $q=0.8$. With the increase of $q$, more orbits can be chaotic.
        }
     \label{Fig9}}
\end{figure*}

\begin{figure*}[htbp]
    \center{
        \includegraphics[scale=0.3]{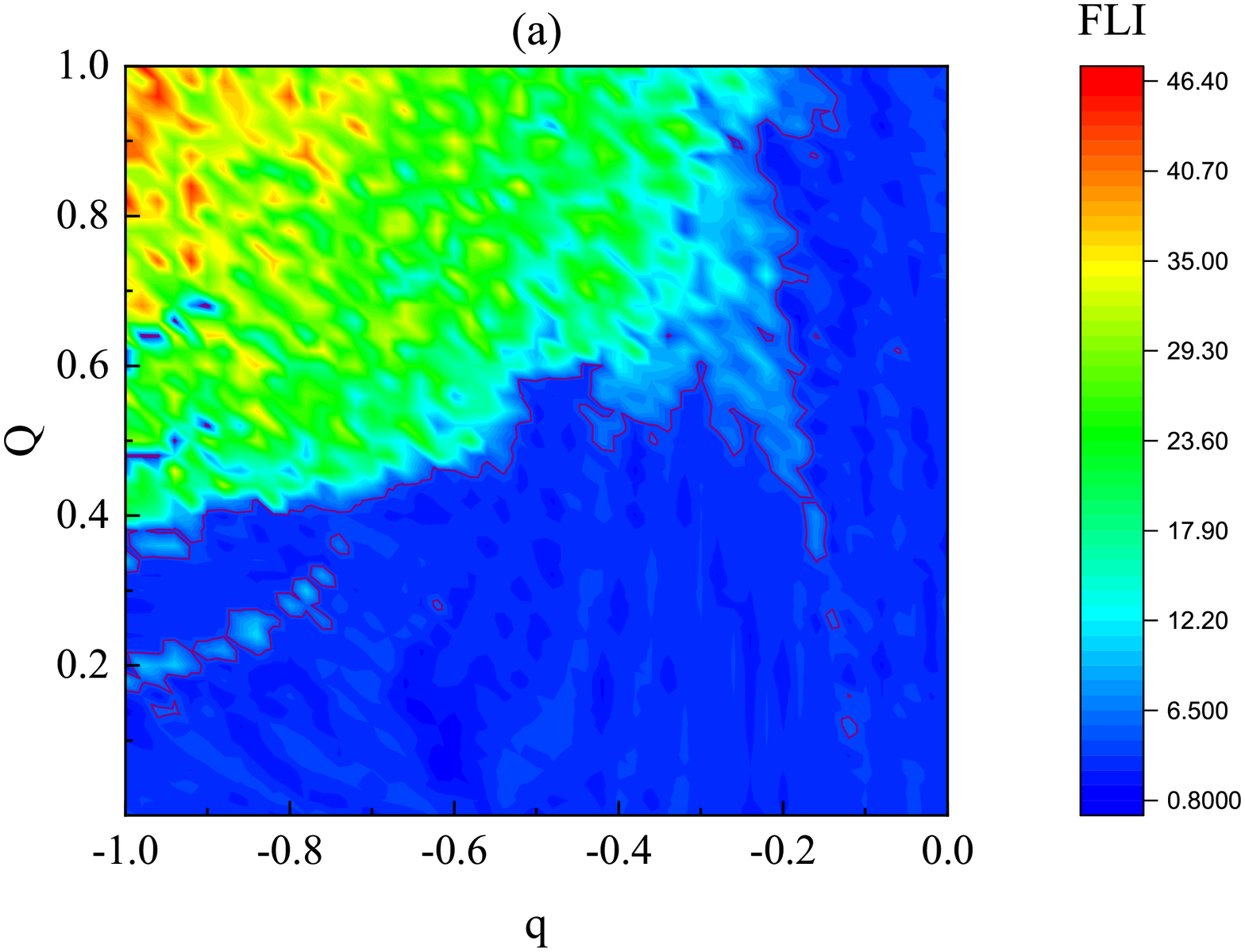}
        \includegraphics[scale=0.3]{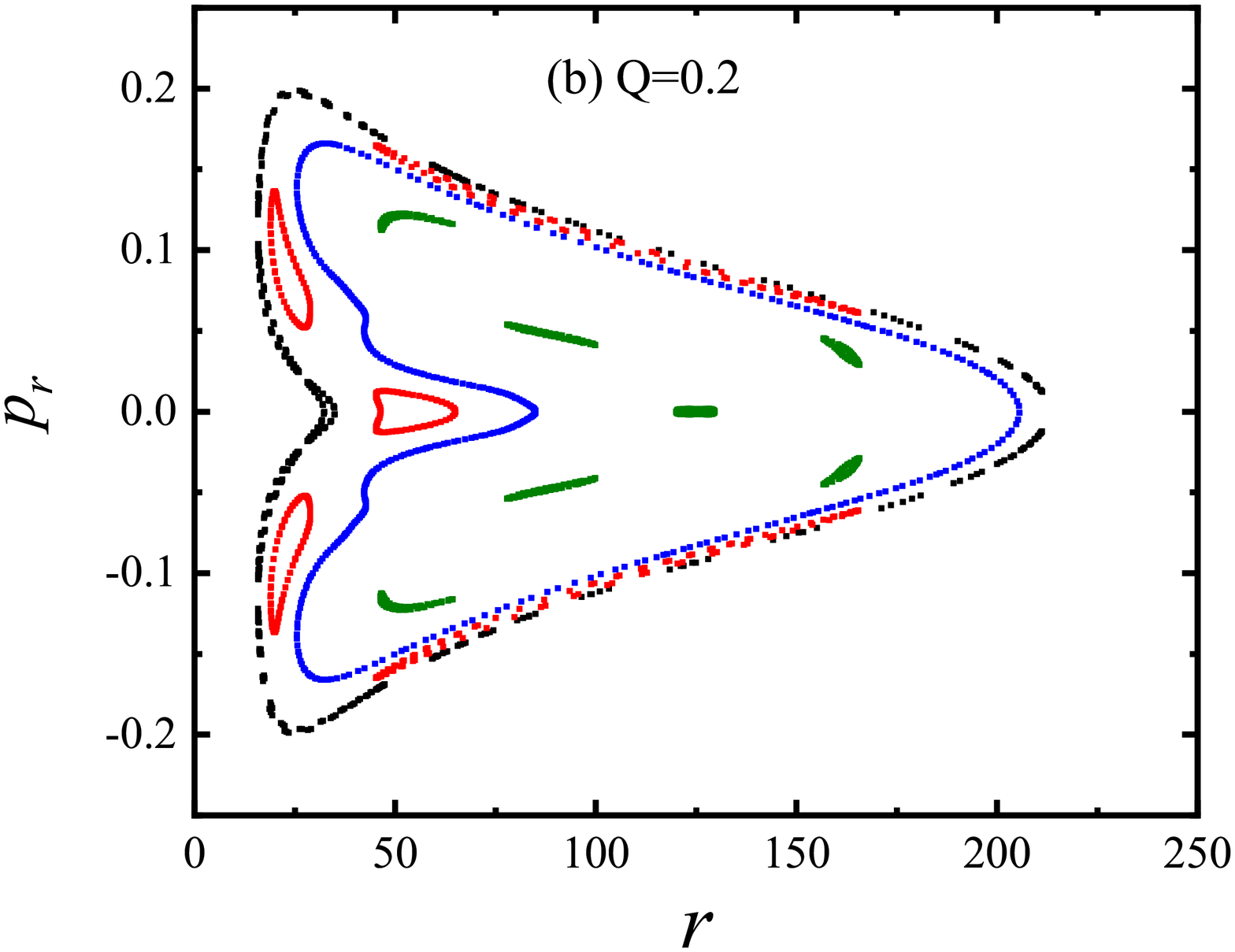}
        \includegraphics[scale=0.3]{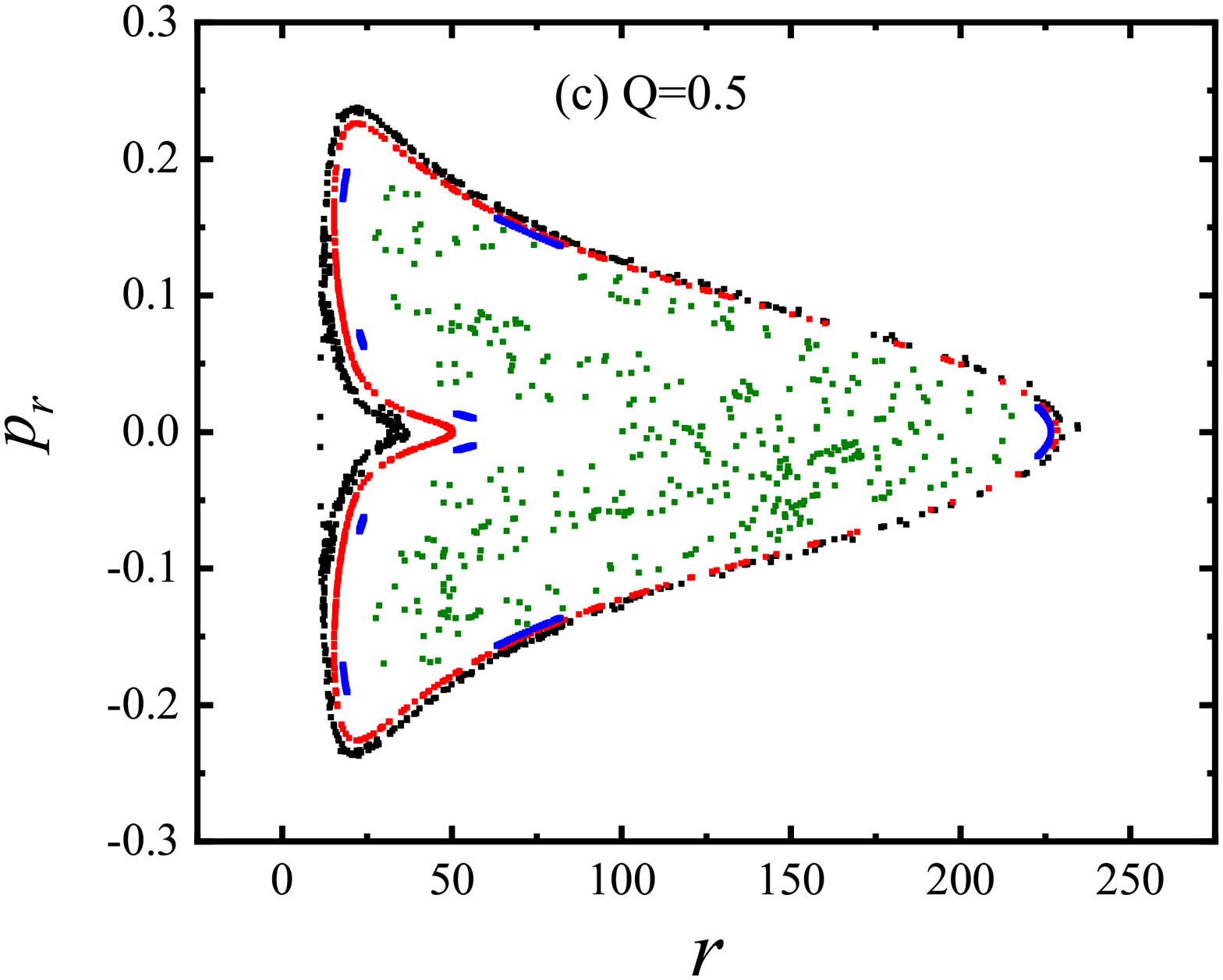}
        \includegraphics[scale=0.3]{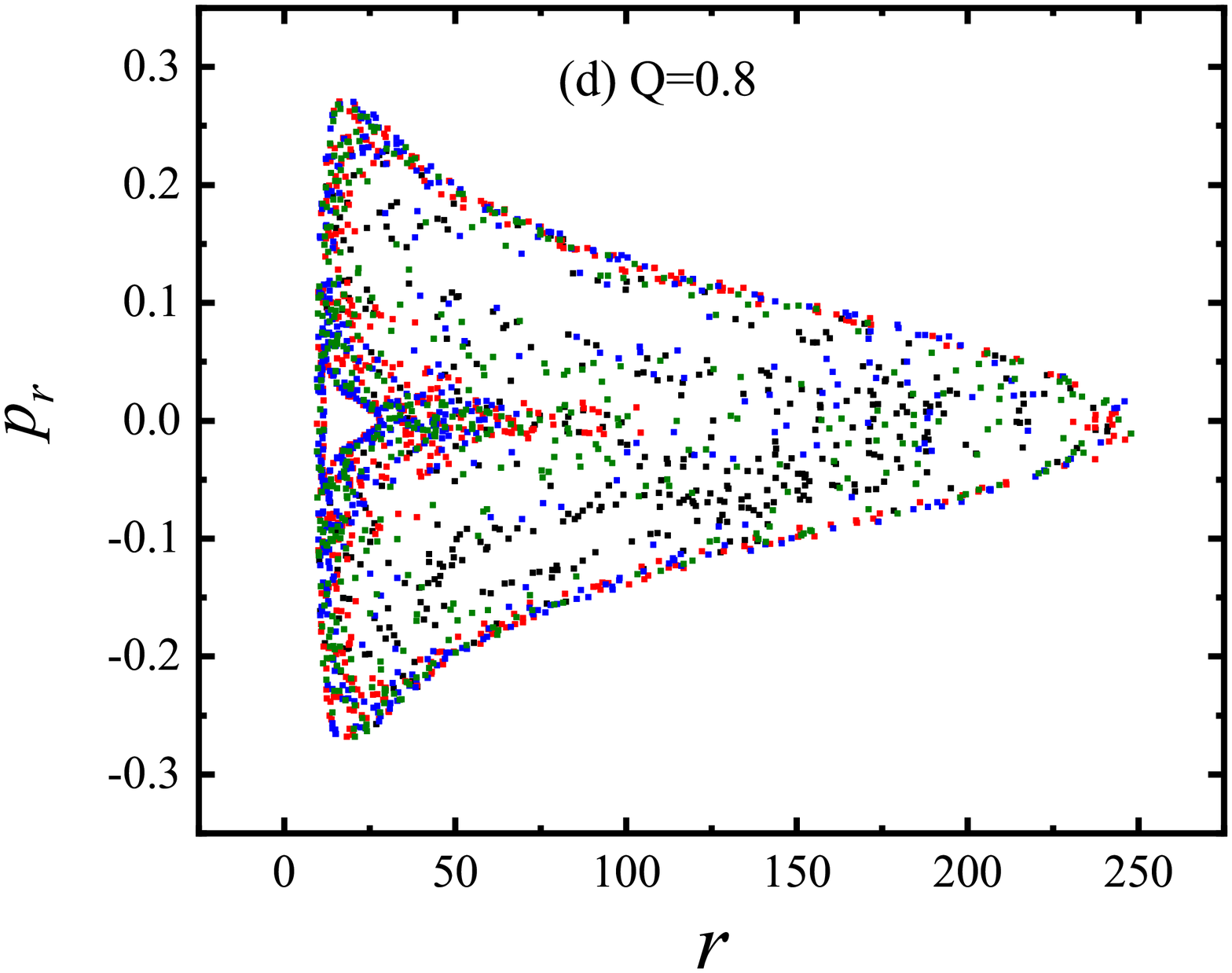}
        \caption{
(a) FLIs for the particle negative charges $q$ and the black hole
positive charges $Q$.  The  other parameters and the initial
separation are the same as those in Fig. 9. (b-d) Poincar\'{e}
sections. The particle charge is $q=-0.5$. The black hole positive
charges are (b) $Q=0.2$,  (c) $Q=0.5$, and  (d) $Q=0.8$. With the
increase of $Q$, more chaotic orbits appear.
  }
     \label{Fig10}}
\end{figure*}

\begin{figure*}[htbp]
    \center{
        \includegraphics[scale=0.2]{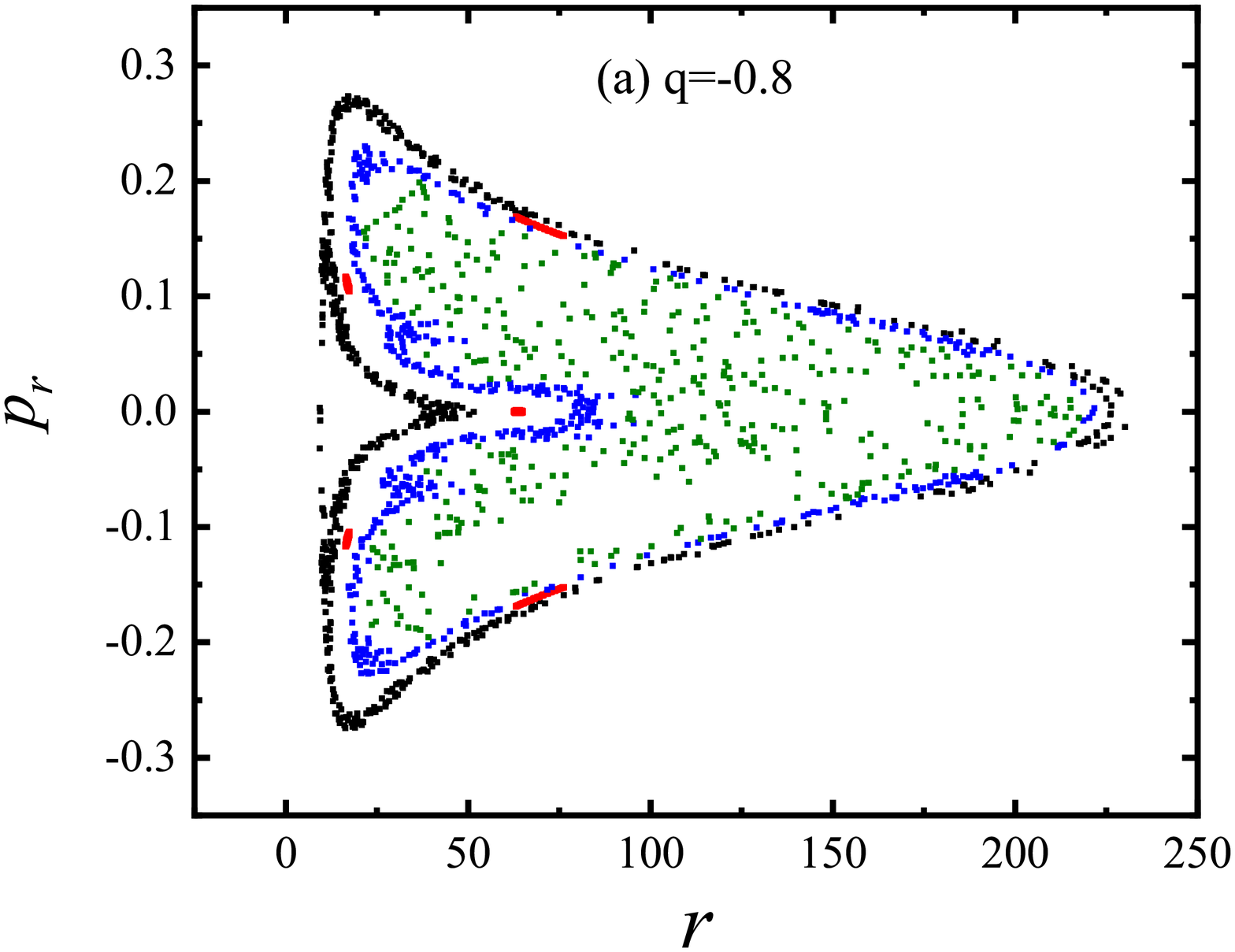}
        \includegraphics[scale=0.2]{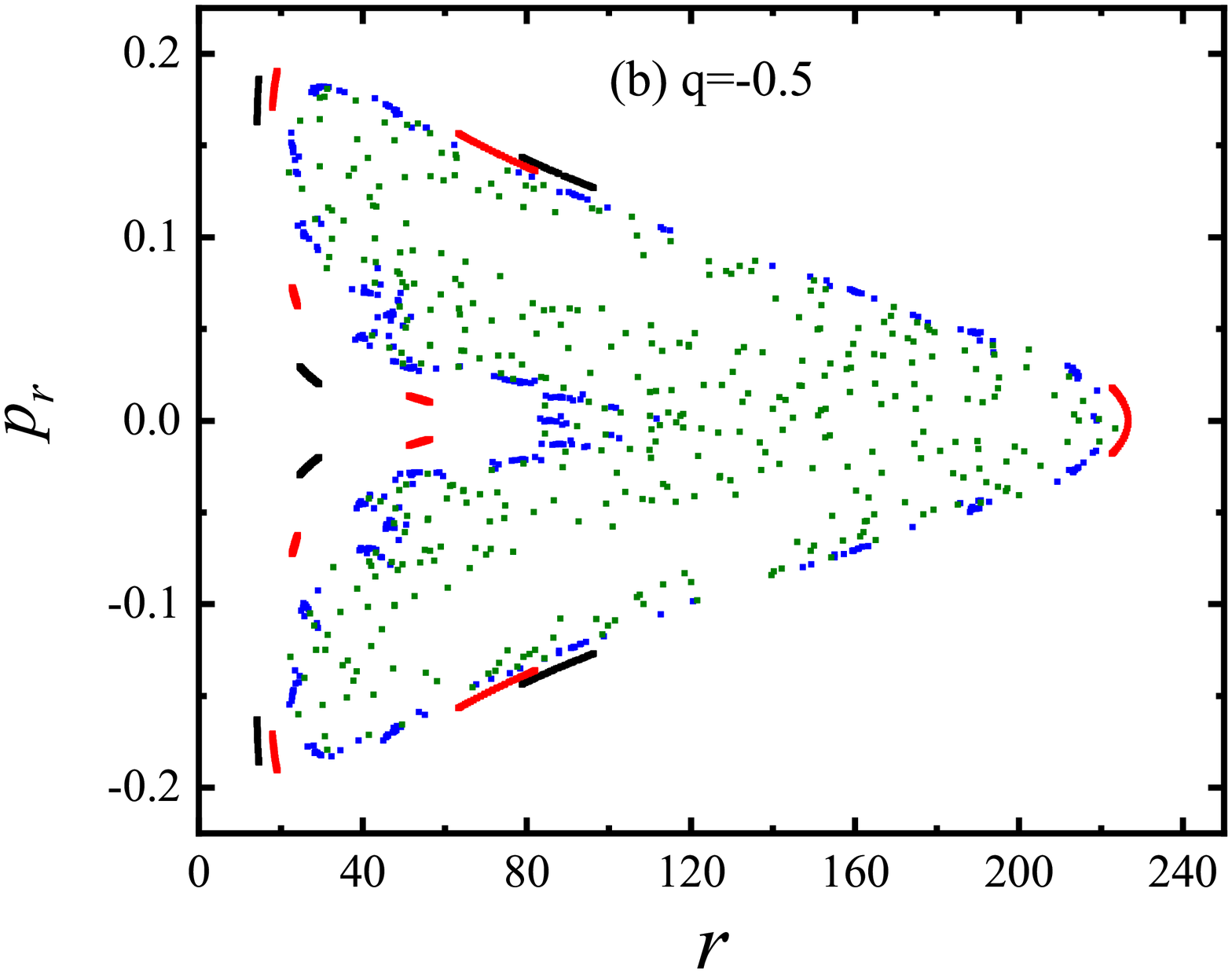}
        \includegraphics[scale=0.2]{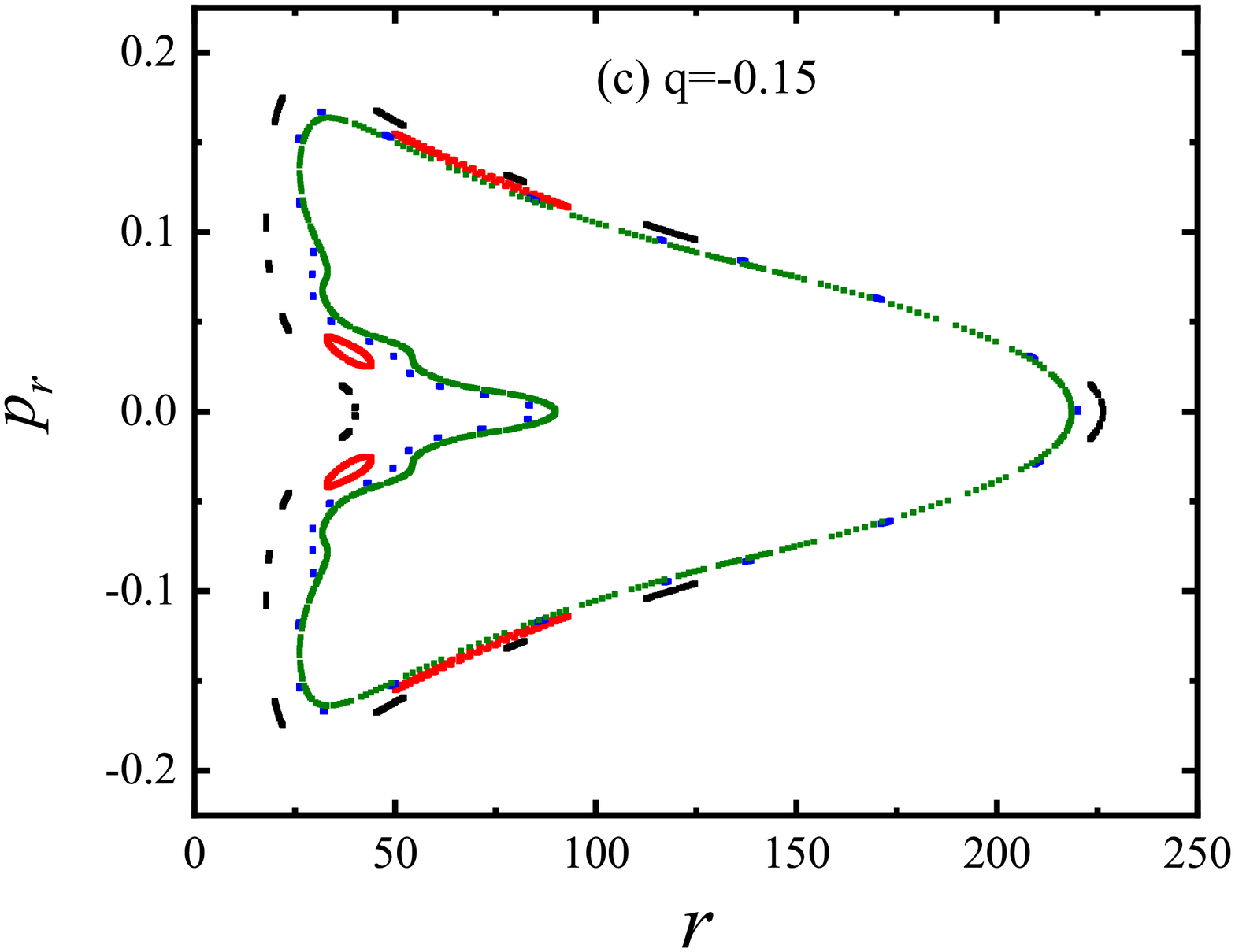}
        \caption{
Poincar\'{e} sections. The black hole positive charge is $Q=0.5$,
and the particle negative charges are (a) $q=-0.8$, (b) $q=-0.5$
and (c) $q=-0.15$. The other parameters are the same as those of
Fig. 10. Chaos easily occurs with the magnitude of particle
negative charge increasing.
  }
     \label{Fig11}}
\end{figure*}

\end{document}